\documentclass[12pt,twoside,pointlessnumbers,smallheadings]{revtex4}
\usepackage{amsmath}
\usepackage{amssymb}
\usepackage{color}
\usepackage{footmisc}

\usepackage{hangcaption,fancybox,feynmp}
\usepackage{indentfirst}
\usepackage{graphicx}
\usepackage{epsfig,subfigure,psfrag}
\usepackage{CJK}
\usepackage{mathrsfs}
\usepackage{float}

\begin{document}


\title{
Discovering extra Higgs boson via pair production of the SM-like Higgs bosons

}
\author{Jia Liu $^{1,2}$, Xiao-Ping Wang $^{1}$ and Shou-hua
Zhu$^{1,3}$}

\affiliation{ $ ^1$ Institute of Theoretical Physics $\&$ State Key
Laboratory of Nuclear Physics and Technology, Peking University,
Beijing 100871, China \\
$ ^2$ Center for Cosmology and Particle Physics, Department of Physics, New York University, New York, NY 10003.\\
$ ^3$ Center for High Energy Physics, Peking University, Beijing 100871, China }


\maketitle

\begin{center}

{\bf Abstract}

\begin{minipage}{15cm}
{\small  \hskip 0.25cm

In the standard model (SM), pair production rate of Higgs boson at the Large Hadron Collider (LHC) is quite low. One usually think that it is extremely important for the measurement of triple Higgs coupling at the high luminosity
LHC.  In this paper, we propose to search for the extra Higgs boson (denoted as $S$) utilizing pair production of the SM-like Higgs boson ($H$) which was discovered in July, 2012.The pair production of $H$ can be huge due to the resonant production of heavy scalar $S$, namely $PP \rightarrow S \rightarrow HH$. The couplings of $H$ with weak gauge boson are similar to ones in the SM and it implies that the couplings between $S$ and gauge bosons are likely suppressed. Provided that $S$ is heavy enough, the decay into weak gauge bosons may not be the dominant modes. Instead $S$ can decay into a pair of $H$ and offer the promising channel to discover it.
In this paper, we studied the 5 promising decay modes of $H$, i.e. $b\bar{b}$, $WW^*$, $ZZ^*$, $\gamma\gamma$ and $\tau^+\tau^-$, and simulated the signals and backgrounds for the 15 combination modes for $HH$ at the LHC with $\sqrt{s}= 14$ TeV and integrated luminosity $\mathcal{L}=1000fb^{-1}$. We found that with the help of suitable selection rules, very good signal to background ratio $S/B$ can be archived in many decay channels, for example $b\bar b+ (WW^*, ZZ^*, \gamma\gamma, \tau^+\tau^-)$, $WW^*+(WW^*, ZZ^*, \gamma\gamma, \tau^+\tau^-)$ and $\tau^+\tau^- +(\gamma\gamma, \tau^+\tau^-)$. For the detailed
 results please refer to Table \ref{tab:discovery} in the text. On the contrary, $b\bar b b\bar b$ mode is less important due to the huge QCD background. However if one has excellent control on light jet mis-tagging, the $b\bar b b\bar b$ mode can be promising to discover
the extra Higgs boson due to its the largest branching ratio compared to other modes.

}

\end{minipage}
\end{center}


\newpage

\section{Introduction\label{introduction}}

One new particle has been discovered at LHC by both ATLAS and CMS collaboration \cite{Aad:2012tfa, Chatrchyan:2012ufa}. The new particle is about $125$GeV, which was first found in the $\gamma \gamma $ decay channel.
Besides to measure the properties of the new particle as precise as possible,  the discovery of the new particle also opens interesting topics, like the possibility of Higgs coupling to dark matter \cite{Zhu:2007gb, Zhu:2005hv, Patt:2006fw, Greljo:2013wja, LopezHonorez:2012kv, Kamenik:2012hn, Djouadi:2011aa} etc.
The couplings among the newly discovered particle and the usual particles in the standard model (SM) have been measured via various decay modes, and one found that they are consistent with those of the SM \cite{ATLAS-CONF-2013-034, CMS-PAS-HIG-13-005, Baglio:2012np, Batell:2012ca, Ellis:2013lra}. Hereafter the new particle is dubbed as SM-like Higgs boson.
However there is still plentiful room for physics beyond the SM (BSM) \cite{Carmi:2012in, Li:2012ku, Bonnet:2012nm, Bonnet:2011yx, ArkaniHamed:2002qy, Agashe:2004rs, LaRochelle:2013zpa, Einhorn:2013tja, Berg:2012cg}.

In the SM, Higgs boson mass is a free parameter. However, provided that the SM is superseded by underlying dynamics at TeV or higher scale, the mass of Higgs boson tends to be close to that higher scale. In a sense, the Higgs boson mass is too light. There are popular solutions to this issue, for example the supersymmetry and little Higgs models. However they are suffered from various difficulties, namely the predicted new companion particles have no sign
at high energy collider. Recently, we proposed a new solution and pointed out the intimate connection between the lightness of the Higgs boson and the spontaneous CP violation \cite{Zhu:2012yv,Hu:2013cda}.  Anyhow there are strong
motivation for searching extra Higgs bosons, denoted as $S$ scalar in this paper.

The couplings of SM-like Higgs boson with weak gauge boson are similar to ones in the SM and it implies that the couplings between the extra Higgs boson and gauge bosons are likely suppressed. Even the extra Higgs boson is heavy enough, the decay into weak gauge boson may not be the dominant modes. Instead S can decay into a pair of SM-like Higgs bosons and offer the promising channel to discover it.
In literature there are many related studies \cite{Harlander:2013mla, Ellwanger:2013ova, Ghosh:2013qga, Kribs:2012kz, Dolan:2012rv, Contino:2012xk, Grober:2010yv, Asakawa:2010xj, Moretti:2010kc, LopezVal:2009qy, Ma:2009kj, Hodgkinson:2009uj, Lafaye:2009vr, Arhrib:2009hc, Kanemura:2008ub, Li:2007bz, Moretti:2007ca, Pierce:2006dh, Binoth:2006ym, Plehn:2005nk,  Dib:2005re, Moretti:2004wa, Baur:2003gp, Baur:2002rb, Baur:2002qd, Djouadi:1999rca, Djouadi:1999gv, Belyaev:1999mx, Plehn:1996wb, Boudjema:1995cb, Jikia:1992mt, Barger:1991jn, Hagiwara:1989xx, Barger:1988kb, Ahriche:2013vqa, Dawson:2012di, Dawson:2012mk, Papaefstathiou:2012qe}. Note that in the SM, there are gluon gluon production of two Higgs through triangle and box diagrams, but the cross-section is quite small, about $30fb$ at LHC14 \cite{Goertz:2013eka, Goertz:2013kp, Shao:2013bz, Papaefstathiou:2012qe, Dawson:2006dm, Jin:2005gw, BarrientosBendezu:2001di, Dawson:1998py, Glover:1987nx}. A recent NNLO calculation for Higgs pair production suggests the cross-section to be $40fb$ \cite{deFlorian:2013jea}. Such small cross-section implies limited statistics in the measurement \cite{Baur:2003gp, Dolan:2012rv, Dolan:2012ac}, which usually need the boosted property in the two Higgs production to cut down the background \cite{Butterworth:2008iy, Butterworth:2008tr}.

In this paper, we will study resonant production of $S$ which subsequently decays into two Higgs ($H$). This process can naturally present in the two Higgs doublet model (2HDM) at small $\tan(\beta)$\cite{Dolan:2012ac}. With the help of resonance production of $S$ and dominant decay branching ratio (BR) to $HH$, the cross-section can be much higher comparing with two Higgs production in SM. We analyze this process at parton level through different combination of SM-like Higgs decay channels with Madgraph 5~\cite{Alwall:2011uj}. The discovery potential for different combination of decay channels are listed in the Table.\ref{tab:discovery}. Most of the combinations have good discovery potential, except the four $b$ quark and $\gamma \gamma ZZ^{*}$ final states. The former does not have large enough $S/B$, while the latter suffers from the low statistics.

 \begin{table}[tbp]
\centering
\begin{tabular}{lccccc}
\hline
decay mode &$b\bar{b}$ &WW &ZZ&$\gamma\gamma$&$\tau\tau$\\ \hline
$b\bar{b}$ &  low $\sigma$  & $ \surd $ & $ \surd $ & $ \surd $ & $ \surd $\\
WW& - & $ \surd $ & $ \surd $  & $ \surd $ & $ \surd $\\
ZZ& - & - & $ \surd $, low N  & low N & $ \surd$, low N \\
$\gamma\gamma$&  - & - & - &  $ \surd$ , low N & $ \surd $\\
$\tau\tau$& - & - & - & - & $ \surd $\\\hline
\end{tabular}
\caption{The discovery potential for different combination of decay channels of Higgs. ``$ \surd $" stands for ``excellent" in searching for the extra Higgs boson. The ``low $\sigma$" means low $S/B$ ratio, while the ``low N" means low signal statistics (signal event number smaller than 100).}\label{tab:discovery}
\end{table}

We arrange this paper as following. We begin with a discussion of the effective model we used for the resonant production of two Higgs in Sec.\ref{sec:model}. Then we explore the final states classified by different Higgs decay combinations and estimate the signal to background ratio ($S/B$) in the Sec.\ref{sec:hh-finalstates}. Sec.\ref{sec:conclusion} contains our discussions and conclusions.

\section{The effective model }\label{sec:model}

For simplicity and as model independent as possible, we do not use 2HDM but start with an effective model and concentrate on the various final states. We assume the effective Lagrangian in the following form:

\begin{eqnarray}
\mathcal{L}=f_1\frac{\sqrt{2}\alpha_s}{12\pi v}SG^a_{\mu\nu}G^{a\mu\nu}+f_2\frac{(m_H)^2}{v}SHH
\end{eqnarray}

\begin{figure}[ht]
\includegraphics[width=15cm]{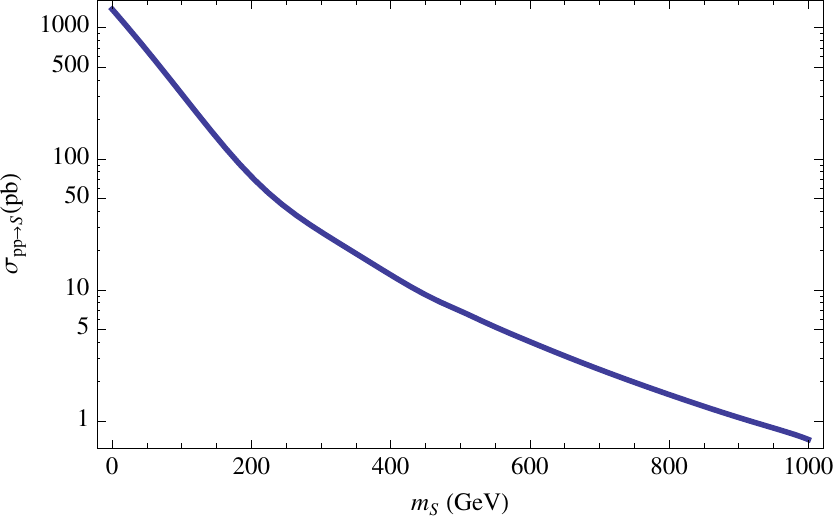}
\caption{The resonant cross-section for $S$.}\label{fig:respro}
\end{figure}

The $S$ can couple to top quark and other color particles in the loop, which results in the coupling to gluon gluon. The $S$ couple to Higgs particle through a portal-like potential, and we assume this coupling provides the significant $S$ decay BR to Higgs, comparing with other decay channels. In order to test the potential of different decay modes of Higgs, we calculate the production cross-section of $pp\rightarrow S\rightarrow HH$ at the LHC with $\sqrt{s}=14$TeV. We assume $f_1 = f_2 = 1$ and the mass of the scalar $S$ equals to 400GeV as the benchmark parameters, and the cross-section of resonant production equals to $10pb$. We plot the resonant cross-section for $S$ in Fig.\ref{fig:respro}. In order to get the potential of Higgs different decay mode, we give the different BR of two Higgs final state according to Ref.\cite{Dittmaier:2011ti} in Table.\ref{tab:BR}. There are constraints on the parameter $f_1$ and $f_2$ but generally quite small. For example, we can assume the $S$ coupling to gluon gluon is induced by a vector-like color particle which does not have other SM charges, and this particle is heavy enough that $S$ can not decay to this color particle. In this case, the $S$ particle can only decay to gluon and Higgs, that the constraints on $f_1$ and $f_2$ are quite weak.
If the top quark is running in loop, the $S$ can also decay to top pair and other gauge boson pair. Since the Higgs $H$ is quite SM-like, the tree level couplings between $S$ and gauge boson should be quite small. However, the top loop can induce coupling between $S$ and gauge boson. Such induced coupling has a ratio of about $0.01 \times N^2_{c}$ to the coupling to gluon gluon, mostly because of the difference between the electroweak and strong coupling \cite{Cao:2009ah}. Since the $S$ also decays to $H$ and top quark, the BR to gauge boson will further decrease, which makes our choice of $f_1$ safe from the $ZZ^{*}$ search for the Higgs. For $S$ decay to top pair, since our resonant production channel is much smaller than the $t \bar{t}$ cross-section, there is very little constraint on $f_1$. For $S$ decay to $H$, there are very few constraints on this channel. Thus, our choice of $f_1$ and $f_2$ are very little constrained by current experiments. With $f_2$ around $\sim O(1)$, we can also guarantee significant production for $S\rightarrow HH$.

 \begin{table}[tbp]
\centering
\begin{tabular}{lccccc}
\hline
decay mode &$b\bar{b}$ &WW &ZZ&$\gamma\gamma$&$\tau\tau$\\ \hline
$b\bar{b}$ &  $3.34 \times 10^{-1}$  & $1.25 \times 10^{-1}$&$1.54 \times 10^{-2}$& $1.33\times 10^{-3}$&$3.68\times 10^{-2}$\\
WW& $1.25 \times 10^{-1}$ & $4.67 \times 10^{-2}$ & $5.77 \times 10^{-3}$  & $4.97 \times 10^{-4}$&$1.38\times10^{-2}$\\
ZZ& $1.54 \times 10^{-2}$ & $5.77 \times 10^{-3}$ & $7.13 \times 10^{-4}$  & $6.14 \times 10^{-5}$&$1.7\times 10^{-3}$\\
$\gamma\gamma$&$1.33 \times 10^{-3}$& $4.97 \times 10^{-4}$ & $6.14 \times 10^{-5}$ &  $5.29 \times 10^{-6}$ &$1.46\times 10^{-4}$\\
$\tau\tau$&$3.68\times 10^{-2}$&$1.38\times10^{-2}$&$1.7\times 10^{-3}$&$1.46\times 10^{-4}$&$4.1\times10^{-3}$\\\hline
\end{tabular}
\caption{The product of two branching ratio of different decay modes for Higgs.}\label{tab:BR}
\end{table}

In our analysis, we assume the luminosity to be $1000fb^{-1}$ for LHC14, which requires the cross-section to be larger than $10^{-6}pb$ to get meaningful statistics. For example, if the two Higgs both decay to $b\bar{b}$, the cross-section is about $3.3pb$, which is large enough to test. If the two Higgs both decay to $W$ bosons, the cross-section is $0.47pb$, while  both decay to $Z$ bosons, the cross-section is $7.1 \times 10^{-3}pb$. If the two higgs boson decay to $\gamma\gamma$, the cross-section is $5.1 \times 10^{-5}pb$ which is fairly small to test. One should note that when the Higgs decay into different final states, there could be an extra factor of two from the different combination of two Higgs decay. This factor will be included in the signal cross-section thereafter, if it is there.

\section{Two Higgs final states}\label{sec:hh-finalstates}
In this section, we will analyze different combinations of two Higgs final states and discuss the signal to background ratio.

\subsection{Two Higgs decay to $b\bar{b} b\bar{b}$}

\begin{figure}[ht]
\includegraphics[width=7.5cm]{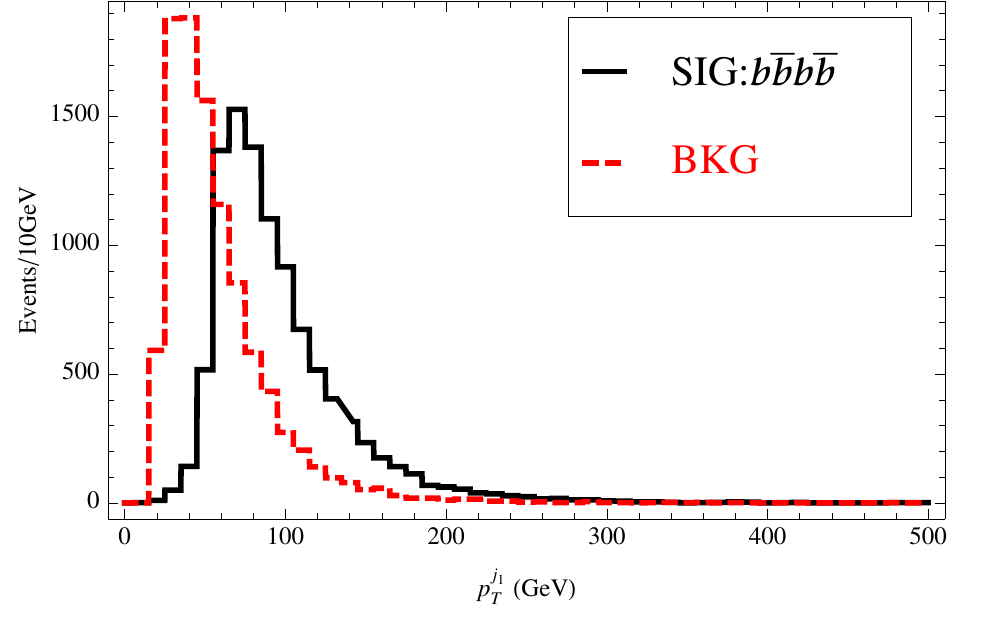}
\includegraphics[width=7.5cm]{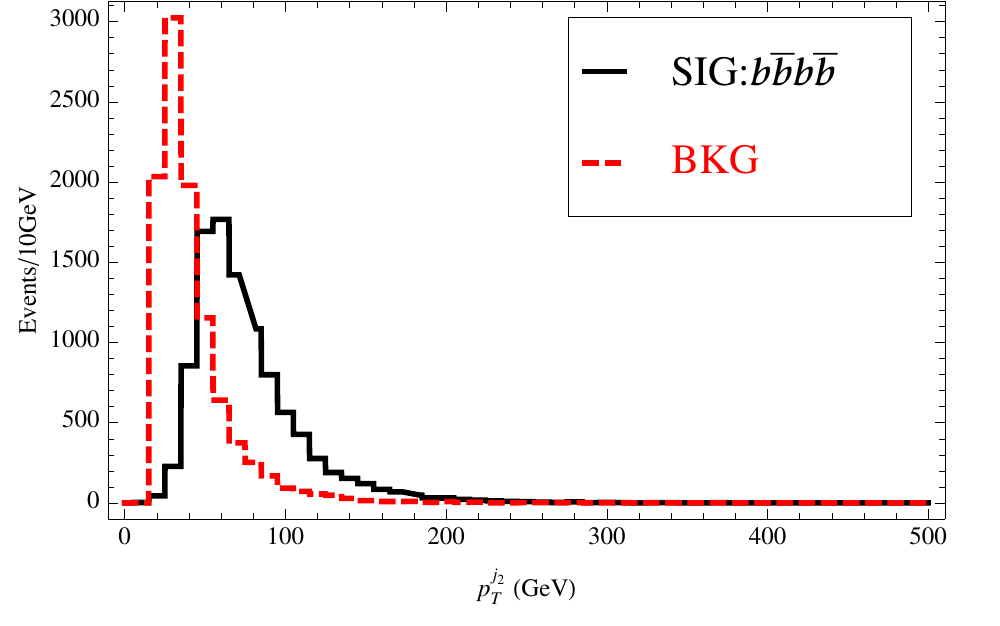}
\includegraphics[width=7.5cm]{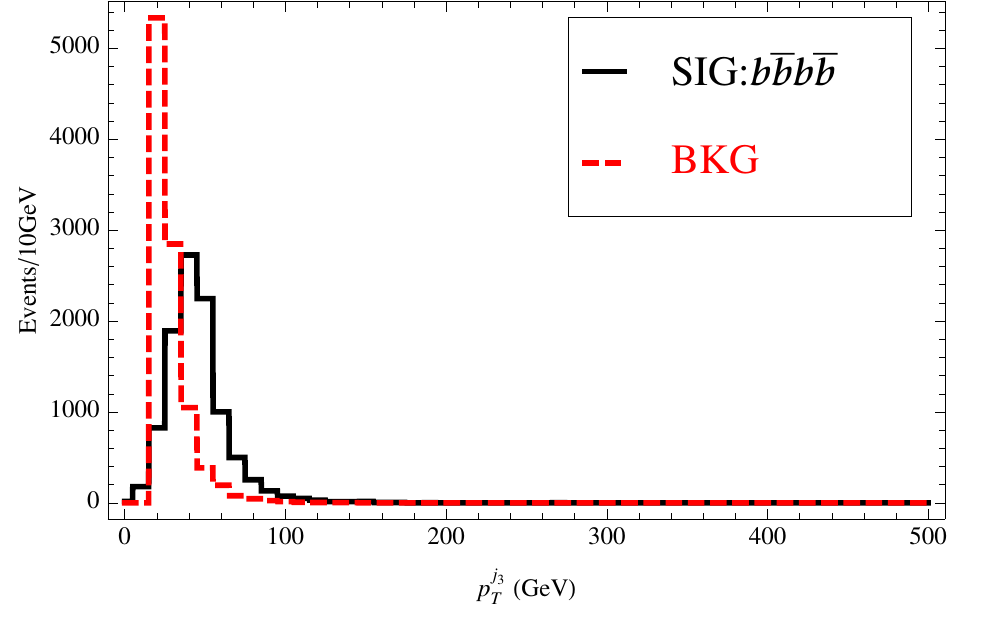}
\includegraphics[width=7.5cm]{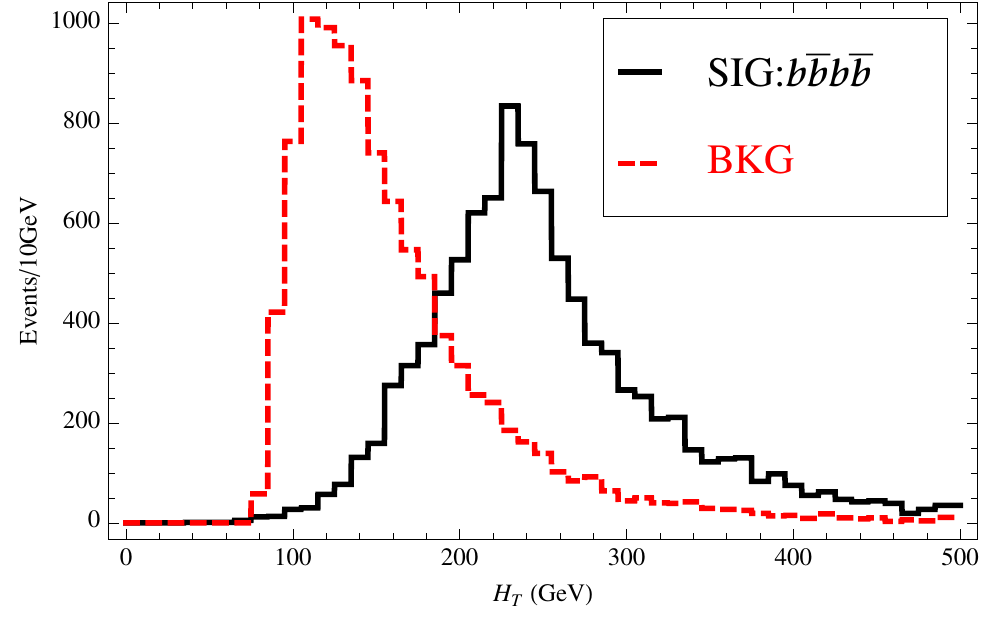}
\caption{The $p_T$ and $H_T$ distribution of signal and background for the $S \rightarrow hh \rightarrow b \bar{b} b \bar{b}$. The dashed red line is SM background and the solid black line is signal.}\label{fig:4b}
\end{figure}

We first begin with the dominant Higgs decay channel $b \bar{b}$, and consider the situation when two Higgs both decay to $b$ quarks. The signal process is $pp\rightarrow S\rightarrow HH\rightarrow(b\bar{b})(b\bar{b})$. At the same time, the dominant SM background contains three processes: $pp\rightarrow bb\bar{b}\bar{b}$, $pp\rightarrow b \bar{b}jj$ and $pp\rightarrow jjjj$, where jets has probability to fake b-jets which described by the b-tag efficiency. For b-quark, the b-tag efficiency is 0.6, while for light flavor jet the mis-tagging is conservatively chosen as 0.02 \cite{Aad:2009wy}.
In order to choose the suitable cut conditions, we give the distribution of $p_T$ and $H_T$ for the jets in the Fig.\ref{fig:4b}.

With the distribution shown in Fig.\ref{fig:4b}, we choose the cut conditions as following:
\begin{eqnarray}
\begin{split}
&p^{j_1}_T\geq 60GeV, p^{j_2}_T\geq 50GeV,  \\
&p^{j_3}_T\geq 40GeV, p^{j_4}_T\geq 20GeV, H_T\geq 180GeV
\end{split}
\end{eqnarray}

Furthermore, for the $b \bar{b}$ coming from the Higgs, the invariant mass of these two b-quark should equal to the Higgs mass. The four b-jets should also reflect the mass of $S$ particle. So we introduce two other cut conditions:

\begin{eqnarray}
\begin{split}
&&m_{b\bar{b}}  \in  [115GeV,135GeV], m_{jjjj}  \in  [360GeV,440GeV]
\end{split}
\end{eqnarray}

After the above cuts, the event number of signal and background are shown in Table.\ref{tab:4b}. The $S/B$ ratio of $b \bar{b} b \bar{b}$ can only reach $4\sigma$ which is not very good even with integrated luminosity of $1000fb^{-1}$. This channel suffers from large QCD background which is quite hard to find the signal. To find the resonant scalar, the two Higgs should have some leptons or photons in the final state to cut down the large QCD background.

 \begin{table}[tbp]
\centering
\begin{tabular}{lcccc}
\hline
           &signal &BKG($b\bar{b}b\bar{b}$)&BKG($b\bar{b}jj$)&BKG(jjjj)\\  \hline
cross-section($pb$)&3.34& 329 & $1.4686\times10^5$ & $2.7536\times10^6$\\
cut efficiency & 0.614 & $6.5\times10^{-4}$ & $1.35\times10^{-3}$ & $2.92 \times 10^{-3}$\\
b-tag&0.1296&0.1296&0.000144&$1.6 \times 10^{-7}$\\
event number &  $2.66\times 10^{5}$ &  $2.77\times10^{4}$ &  $2.85\times 10^{4}$  & 1282 \\
$\frac{S}{B}$ ratio& 4.6 \\  \hline
\end{tabular}
\caption{The cut flow table for the $S \rightarrow HH \rightarrow b \bar{b} b \bar{b}$ signal. The ``b-tag" means b-tag efficiency for b quark and other light quarks. In the cross-section for background, we may have added prior $p_T$ requirement in the Madgraph to generate events more efficiently.}\label{tab:4b}
\end{table}

\subsection{Two Higgs decay to $b\bar{b}+\gamma\gamma$ }
\begin{figure}[ht]
\includegraphics[width=7.5cm]{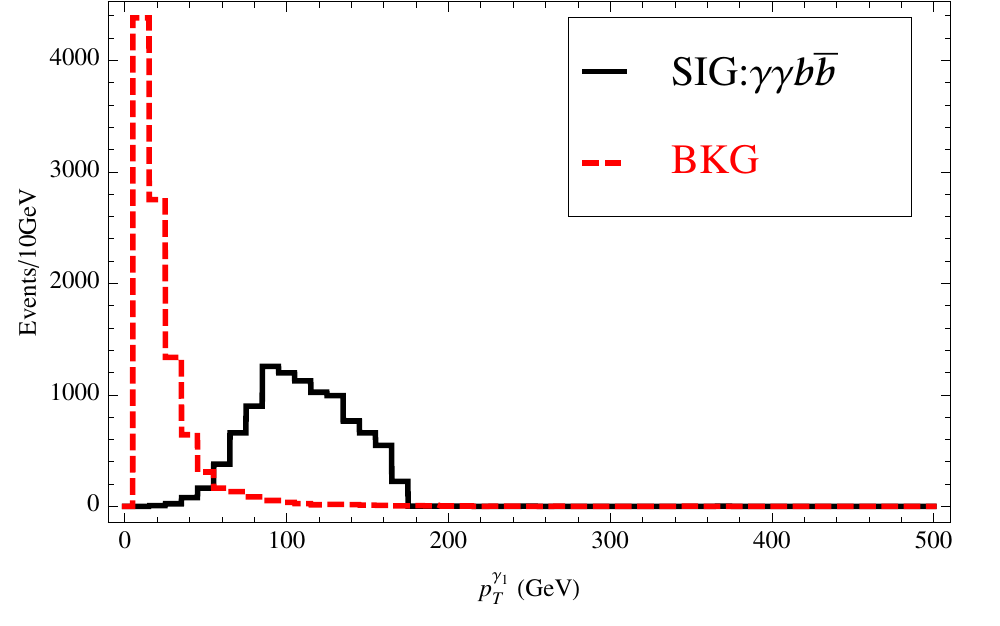}
\includegraphics[width=7.5cm]{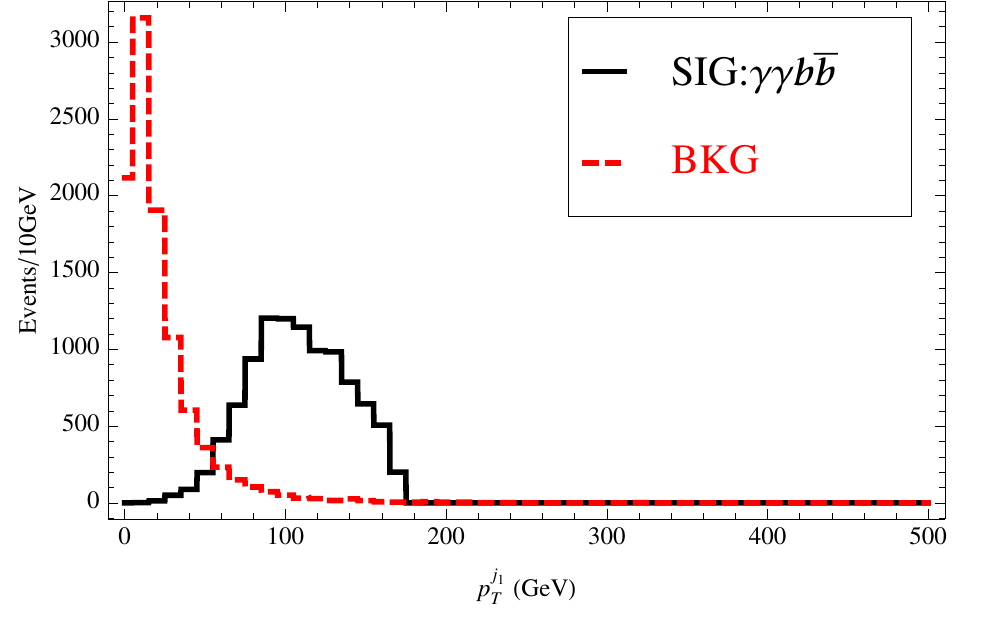}
\includegraphics[width=7.5cm]{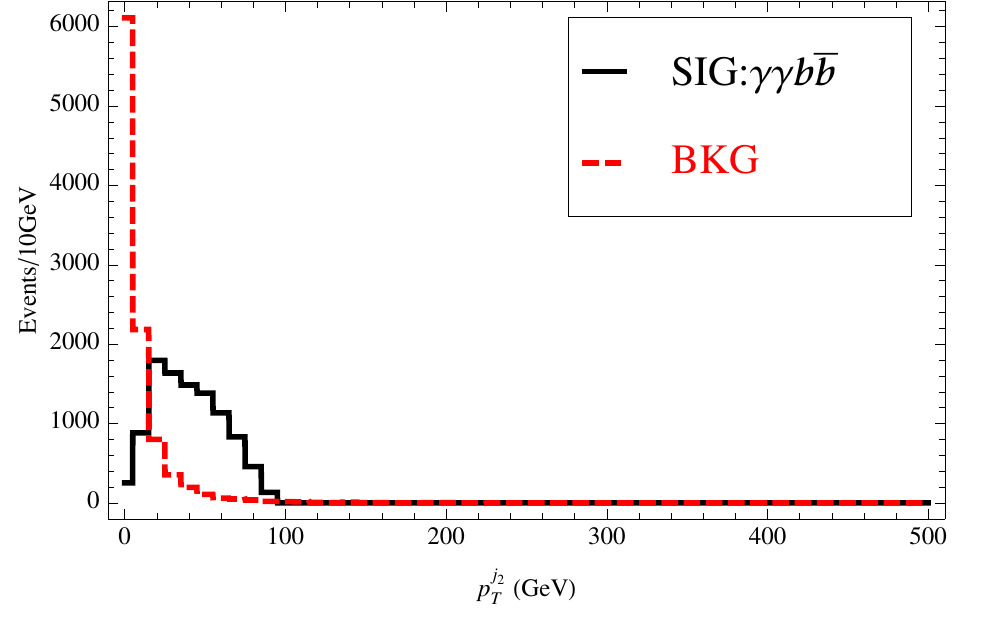}
\includegraphics[width=7.5cm]{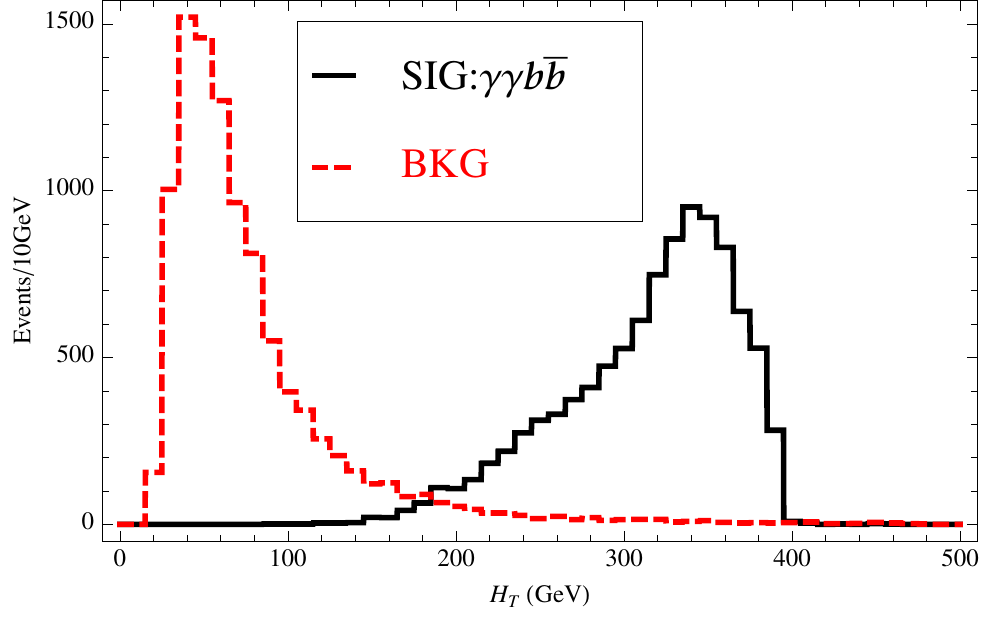}
\caption{The $p_T$ and $H_T$ distribution of signal and background for the $b\bar{b}+\gamma\gamma$ final state. The dashed red line is background and the solid black line is signal.}\label{fig:2a2b}
\end{figure}

In this section, we examine the signal process $pp\rightarrow S\rightarrow HH\rightarrow(\gamma\gamma)(b\bar{b})$. The SM background contains $pp\rightarrow \gamma\gamma b\bar{b}$ and $pp\rightarrow \gamma\gamma jj$ final state. There are other background from jet faking which is hard to simulate at parton level, which we neglect here.  In order to choose the suitable cut conditions, we give the $p_T$ and $H_T$ distribution for the jets and photons in Fig.\ref{fig:2a2b}.

With the distribution shown in Fig.\ref{fig:2a2b}, we choose the cut conditions as following:
\begin{eqnarray}
\begin{split}
&p^{j_1}_T\geq 60GeV, p^{j_2}_T\geq 20GeV\\
&p^{\gamma_1}_T\geq 40GeV, p^{\gamma_2}_T\geq 20GeV,H_T\geq 200GeV
\end{split}
\end{eqnarray}

where the $H_T$ means the $p_T$ sum of photons, jets, leptons and missing energy, if they are there. In this case, since there is no neutrino, lepton and missing energy, the $H_T$ only includes $p_T$ sum of jets and photons. We use this definition of $H_T$ afterwards. Moreover, the $\gamma \gamma$ and the $b\bar{b}$ for signal comes from the Higgs, so the invariant mass of these particles should be equals to the Higgs mass. The invariant mass of two photons and two $b$ quarks also reflect the $S$ mass. Thus, we introduce two cut conditions:

\begin{eqnarray}
\begin{split}
&&m_{\gamma\gamma},  m_{b\bar{b}} \in [115GeV,135GeV], m_{\gamma\gamma b\bar{b}}  \in  [360GeV,440GeV]
\end{split}
\end{eqnarray}

After above cuts, the event number of signal and background for $b\bar{b}+\gamma\gamma$ are shown in Table.\ref{tab:2a2b}. This channel is a very promising for the resonant scalar, due to the large resonant production cross-section. The event number for signal is significantly large and is worth looking at.

 \begin{table}[tbp]
\centering
\begin{tabular}{lccc}
\hline
event number &signal &BKG($\gamma\gamma b\bar{b}$)&BKG($\gamma\gamma jj$)\\ \hline
cross-section($pb$)&0.0266& 0.11195 &120\\
cut efficiency & 0.77 & $1.5\times10^{-4}$ & $6 \times 10^{-4}$  \\
b-tag& 0.36 & 0.36 &0.0004\\
event number & $7.4\times 10^{3}$ & 6  &   29   \\
$\frac{S}{B}$ ratio& 211 \\ \hline
\end{tabular}
\caption{The cut flow table for the final state with one Higgs decay to $\gamma\gamma$ and the other decay to $b\bar{b}$. In the cross-section for background, we may have added prior $p_T$ requirement in the Madgraph to generate events more efficiently.}\label{tab:2a2b}
\end{table}

\subsection{Two Higgs decay to $b\bar{b} + ZZ^{*}$}
\begin{figure}[ht]
\includegraphics[width=7.5cm]{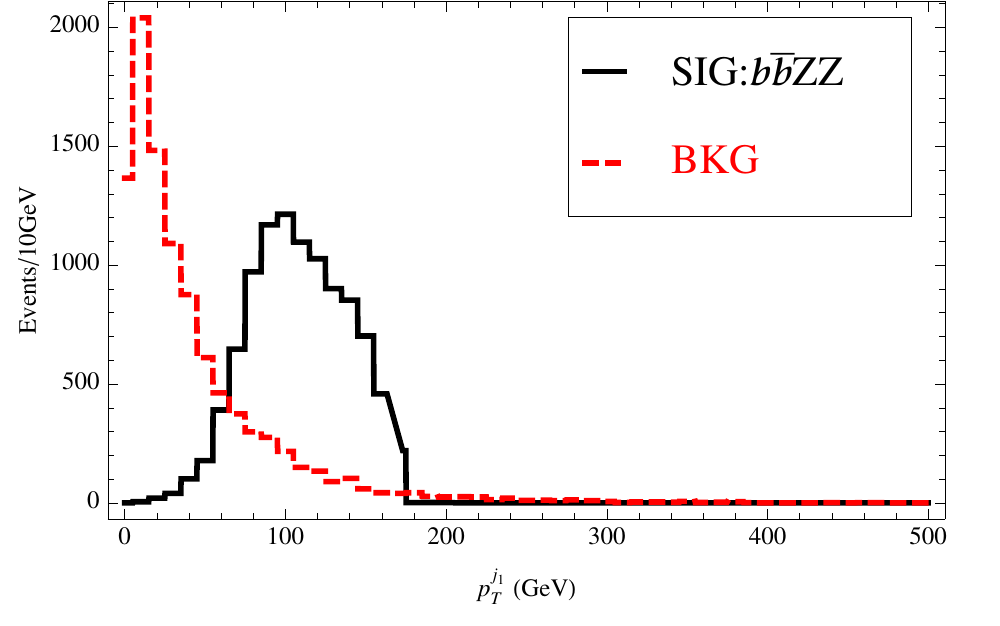}
\includegraphics[width=7.5cm]{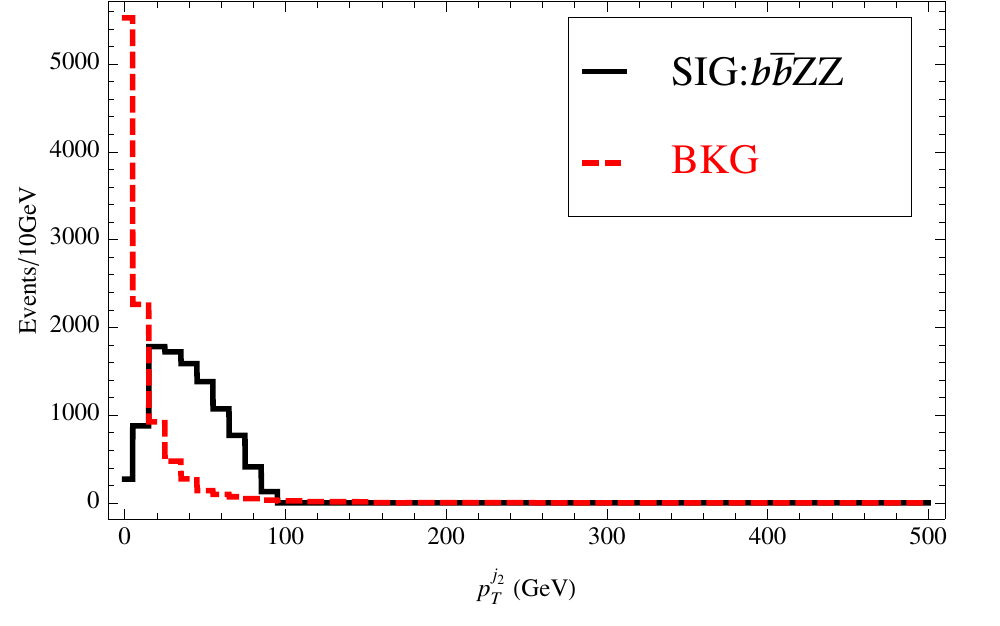}
\includegraphics[width=7.5cm]{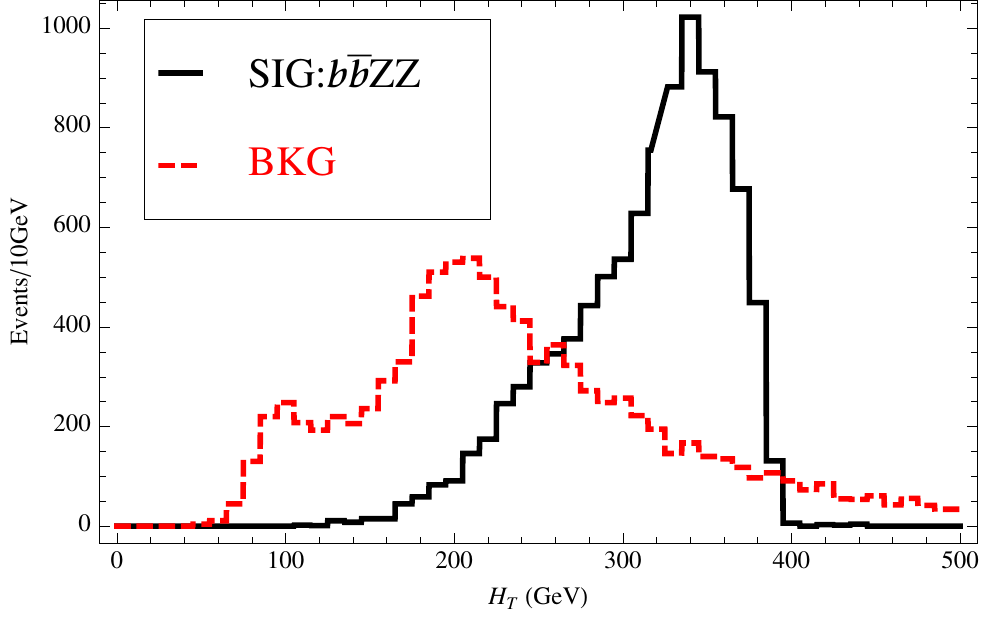}
\caption{The jet $p_T$ and the $H_T$ distribution of signal and background for final state $b\bar{b} + ZZ^{*}$. The dashed red line is background, while the solid black line is signal.}\label{fig:2b2Z}
\end{figure}

In this section, we consider the signal process $pp\rightarrow S\rightarrow HH\rightarrow(b\bar{b})(ZZ^{*}) $ with both Z decay to leptons. At the same time, the SM background contains $pp\rightarrow b\bar{b} Z l^-l^+$ and $pp\rightarrow jj Z l^-l^+$ with Z decay leptonically. In order to choose the suitable cut conditions, we give the distribution of $p_T$ and $H_T$ for the jets in the Fig.\ref{fig:2b2Z}.

With the distribution shown in Fig.\ref{fig:2b2Z}, we choose the cut conditions as following:
\begin{eqnarray}
\begin{split}
&p^{j_1}_T\geq 60GeV, p^{j_2}_T\geq 20GeV\\
&p^{l_1}_T\geq 20GeV, H_T\geq 260GeV
\end{split}
\end{eqnarray}

 We also impose the cuts on invariant mass as following,

\begin{eqnarray}
\begin{split}
&&m_{b\bar{b}} , m_{llll} \in [115GeV,135GeV],  m_{b\bar{b}llll} \in [360GeV,440GeV].
\end{split}
\end{eqnarray}

After above cuts, the event number of signal and background are shown in Table.\ref{tab:2b2Z}. In this channel, the $S/B$ ratio is very good due to the small cross-section of background. The four lepton final states has very small cross-section due to small electroweak coupling comparing with strong coupling. The $b\bar{b}llll$ channel is a very clean and worth looking at.

 \begin{table}[tbp]
\centering
\begin{tabular}{lccc}
\hline
event number &signal &BKG($b\bar{b}llll$)&BKG(jjllll)\\ \hline
cross-section($pb$)&$3.08\times 10^{-3}$&$7.8\times 10^{-4}$&0.013\\
Cut efficiency & 0.611 & 0 & $10^{-4}$ \\
b-tag&0.36&0.36&0.0004\\
Event number &   677   &    0  &   $5.2 \times 10^{-4}$ \\
$\frac{S}{B}$ ratio&$1.3\times 10^{6}$\\ \hline
\end{tabular}
\caption{The cut flow table for two Higgs decay to $b\bar{b} + ZZ^{*}$. The cross-section for signal includes the $Z$ leptonic decay BR.}\label{tab:2b2Z}
\end{table}

\subsection{Two Higgs decay to $b\bar{b} + WW^{*}$}
\begin{figure}[ht]
\includegraphics[width=7.5cm]{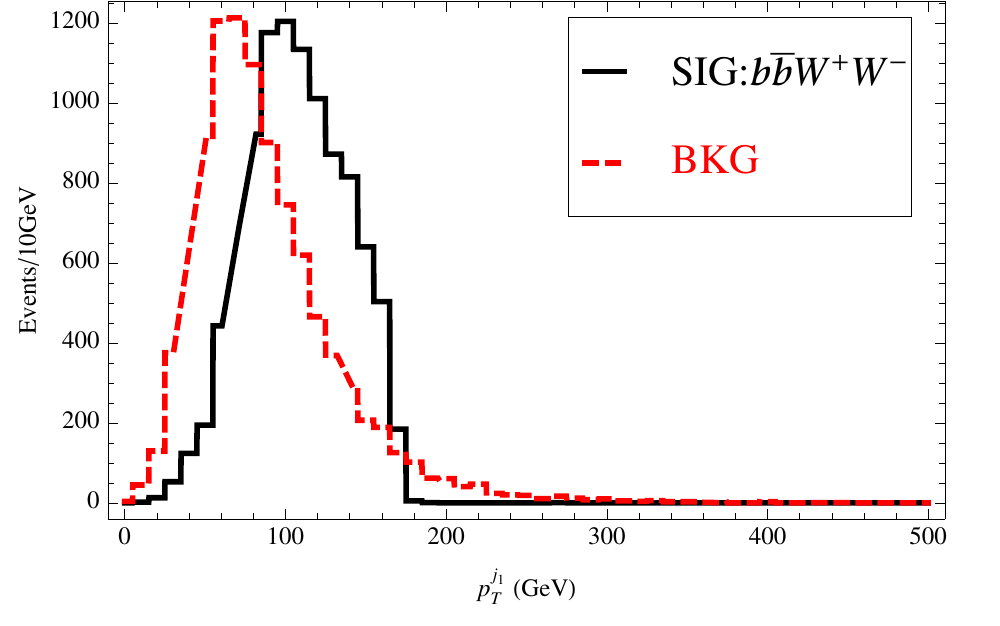}
\includegraphics[width=7.5cm]{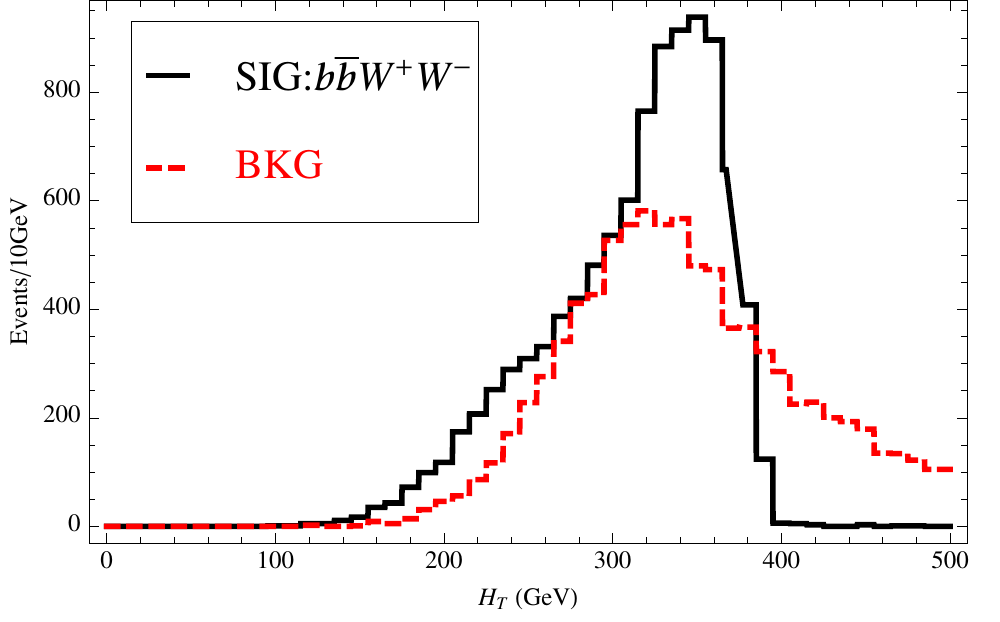}
\caption{The jet $p_T$ and $H_T$ distribution of signal and background for $b\bar{b} + WW^{*}$ final state. The dashed red line is background and the solid black line is signal.}\label{fig:2b2W}
\end{figure}

The signal process we consider is $pp\rightarrow S\rightarrow HH\rightarrow(b\bar{b})(WW^{*})  $  with $W$ decay leptonically. At the same time, the SM background contains $pp\rightarrow b\bar{b}l^-l^+\nu\nu$ and $pp\rightarrow jjl^-l^+\nu\nu$, with the former one includes the top quark pair background. In order to choose the suitable cut conditions, we give the $p_T$ and $H_T$ distribution in the Fig.\ref{fig:2b2W}.

With the distributions shown in Fig.\ref{fig:2b2W}, we choose the cut conditions as follow:
\begin{eqnarray}
p^{j_1}_T\geq 80GeV.
\end{eqnarray}

We also impose the cuts on invariant mass to further suppress the background.

\begin{eqnarray}
\begin{split}
&&m_{b\bar{b}}, m_{ll\nu\nu} \in  [115GeV,135GeV], m_{b\bar{b}ll\nu\nu} \in [360GeV,440GeV]
\end{split}
\end{eqnarray}

After above cuts, the event number of signal and background are shown in Table.\ref{tab:2b2W}. In the above cuts, we have used the momentum of neutrinos which could be reconstructed by requiring the on-shell mass conditions. The backgrounds are mostly cut down by the invariant mass requirement. In this channel top pair background has also been considered, but the cut efficiency is zero, reflecting very small background after such cuts. The $S/B$ ratio of this channel is very good and is promising to be found.

 \begin{table}[tbp]
\centering
\begin{tabular}{lccc}
\hline
    &signal &BKG($b\bar{b}l^-l^+\nu\nu$)&BKG($jjl^-l^+\nu\nu$)\\ \hline
cross-section($pb$)&0.27&21.75&0.8\\
cut efficiency&  0.845 &0&0\\
b-tag&0.36&0.36&0.0004\\
event number& $8.2 \times 10^{4}$ & 0 & 0  \\
$\frac{S}{B}$ ratio&$\infty$\\ \hline
\end{tabular}
\caption{The cut flow table for two Higgs decay to $b\bar{b} + WW^{*}$.}\label{tab:2b2W}
\end{table}

\subsection{Two Higgs decay to $ZZ^{*} +ZZ^{*}$}
\begin{figure}[ht]
\includegraphics[width=7.5cm]{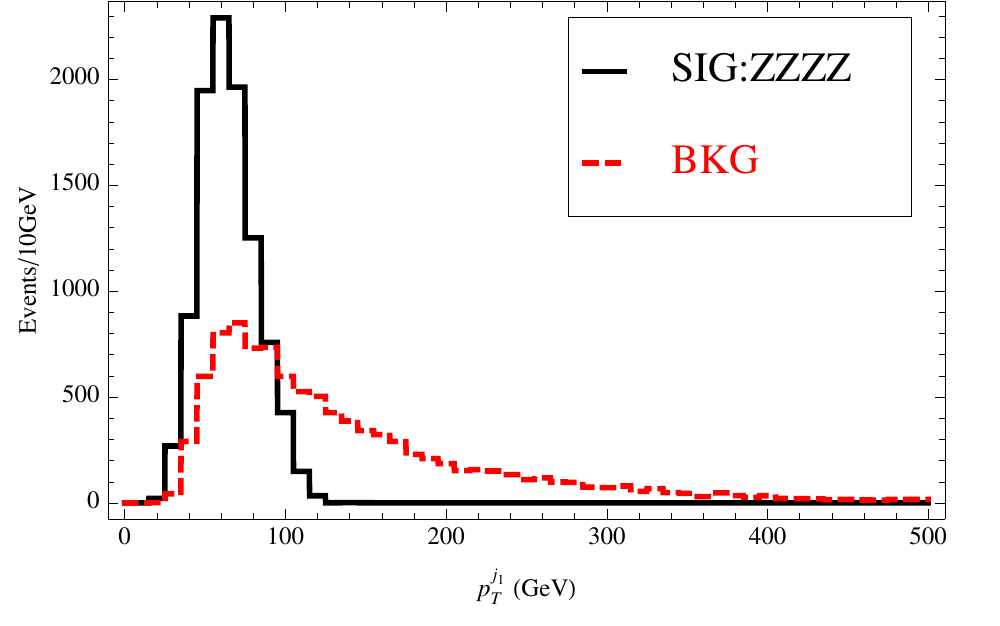}
\includegraphics[width=7.5cm]{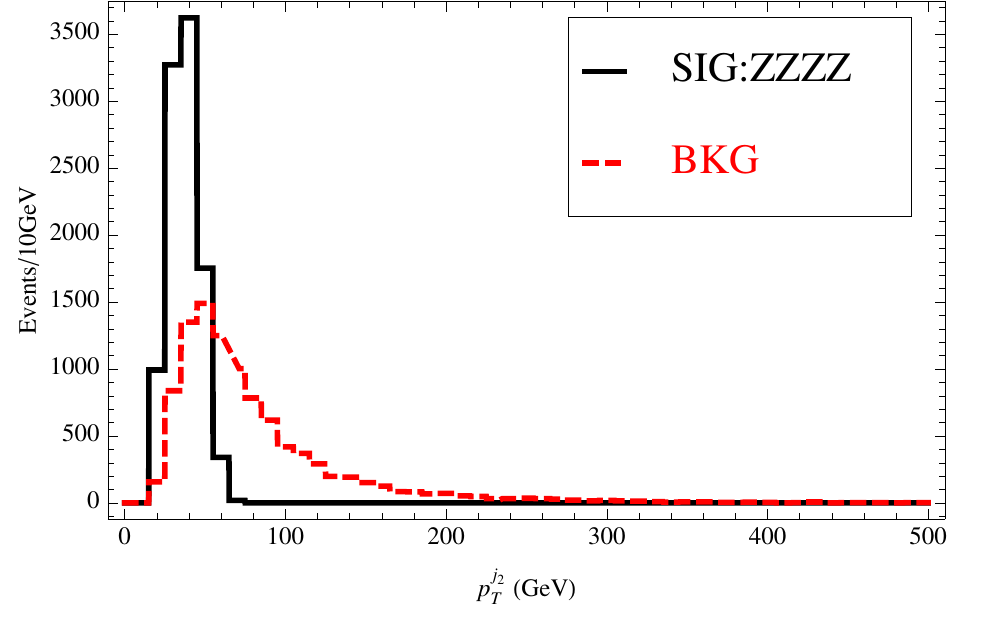}
\includegraphics[width=7.5cm]{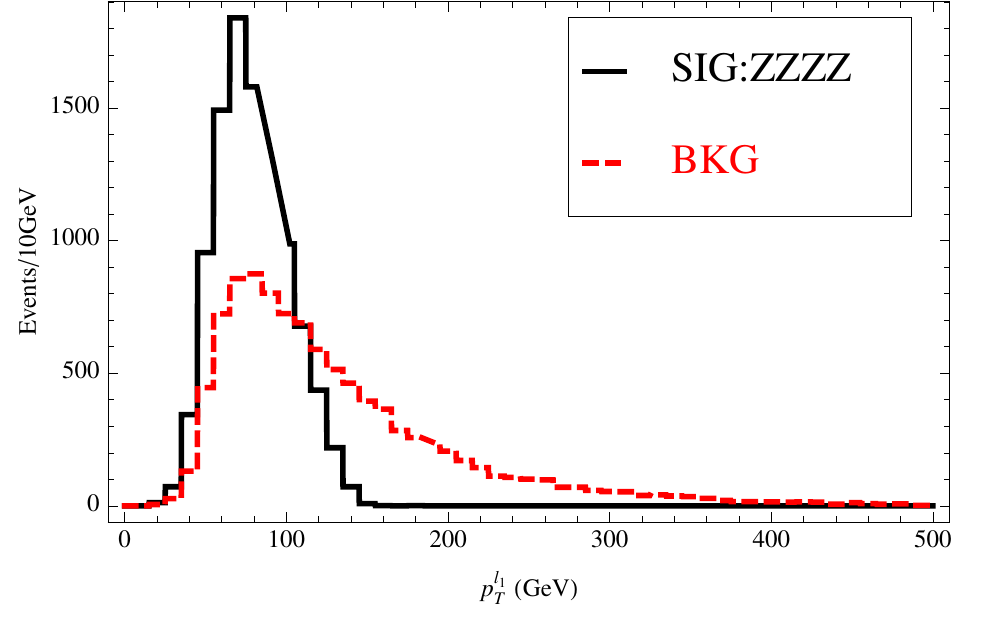}
\includegraphics[width=7.5cm]{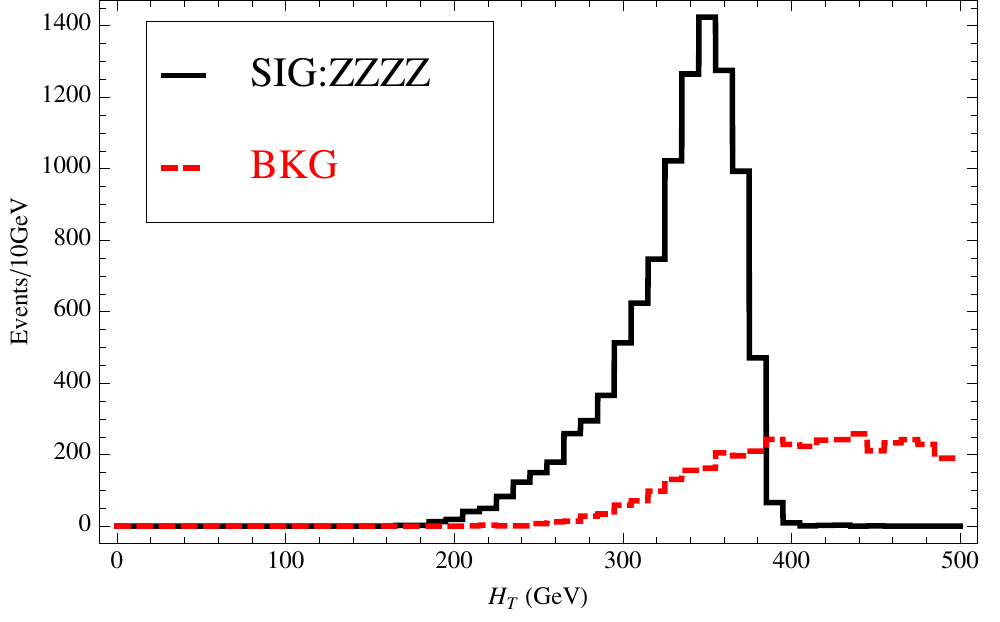}
\caption{The $p_T$ and $H_T$ distribution for signal $ZZ^{*} +ZZ^{*}$ and corresponding background. The dashed red line is background, while the solid black line is signal.}\label{fig:4Z}
\end{figure}

In this section, we assume the signal process $pp\rightarrow S\rightarrow HH\rightarrow(Zl^-l^+)(Zjj)\rightarrow(l^-l^+l^-l^+)(jjjj)$. We arrange the Z bosons from one Higgs decay leptonically, while the other one decays hadronically to get larger signal cross-section. At the same time, the background contains: $pp\rightarrow (Zl^-l^+)(Zjj)\rightarrow (l^-l^+l^-l^+)(jjjj)$. In order to choose the suitable cut conditions, we give the distribution of $p_T$ and $H_T$ in the Fig.\ref{fig:4Z}.

With the distributions shown in Fig.\ref{fig:4Z}, we choose the cut conditions as follow:
\begin{eqnarray}
\begin{split}
&p^{j_1}_T\leq 80GeV, p^{j_2}_T\leq 60GeV\\
&p^{l_1}_T\leq 120GeV, H_T\leq 400GeV
\end{split}
\end{eqnarray}

We also impose the invariant mass cut conditions below,

\begin{eqnarray}
\begin{split}
&&m_{4l}, m_{4j} \in [115GeV,135GeV], m_{4l4j}\in [360GeV,440GeV].
\end{split}
\end{eqnarray}

After above cuts, the event number for signal $ZZ^{*} +ZZ^{*}$ and background are shown in Table.\ref{tab:4Z}. This channel is quite clean even before the cuts, the signal even has larger cross-section due to multi-leptons in the final states. After the cuts, the background vanishes which means very small background. However, the event number of signal is not very large, of about 46 that low statistics makes this channel harder than others even with very clean background.

 \begin{table}[tbp]
\centering
\begin{tabular}{lcc}
\hline
event number &signal &BKG($l^-l^+l^-l^+jjjj$)\\ \hline
cross-section($pb$)&$7.0 \times 10^{-5}$ & $2.24 \times 10^{-5}$\\
cut efficiency&  0.66 &0\\
event number & 46 & 0 \\
$\frac{S}{B}$ ratio  & $\infty$  &  \\  \hline
\end{tabular}
\caption{The cut flow table for $ZZ^{*} +ZZ^{*}$ channel.}\label{tab:4Z}
\end{table}

\subsection{Two Higgs decay to $WW^{*} + WW^{*}$}

\begin{figure}[ht]
\includegraphics[width=7.5cm]{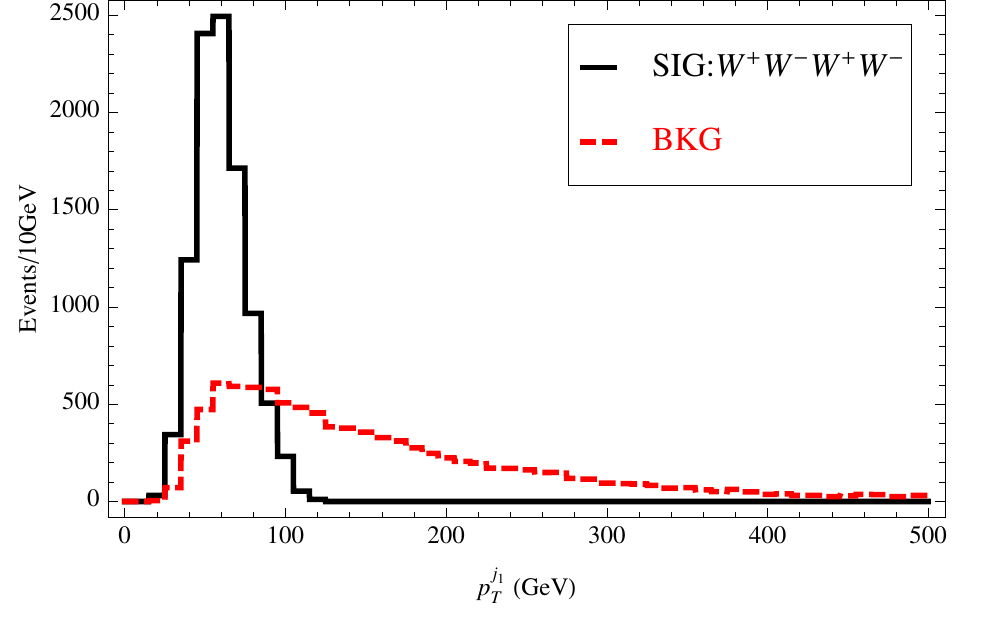}
\includegraphics[width=7.5cm]{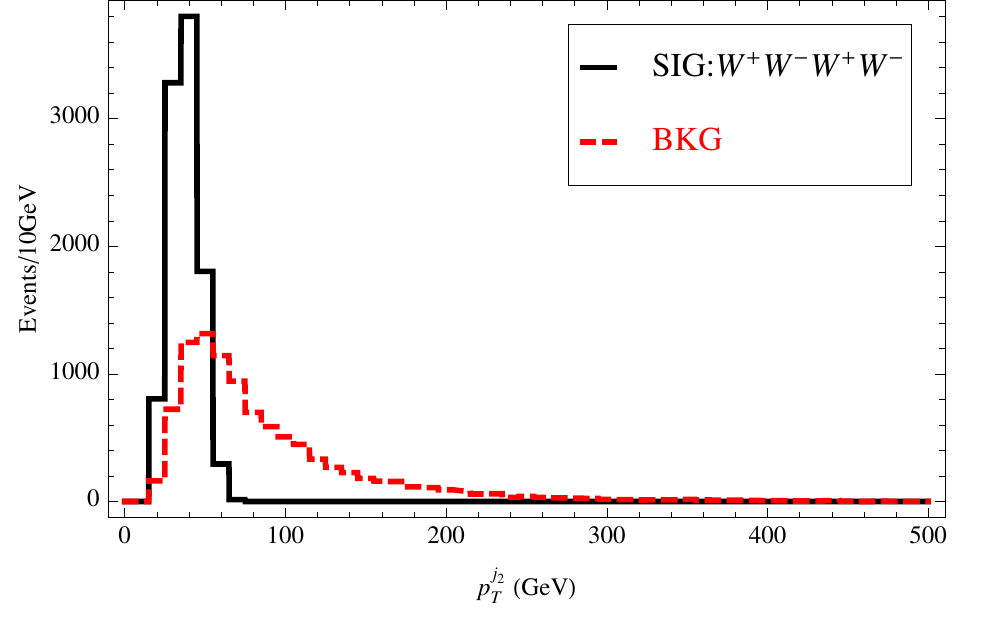}
\includegraphics[width=7.5cm]{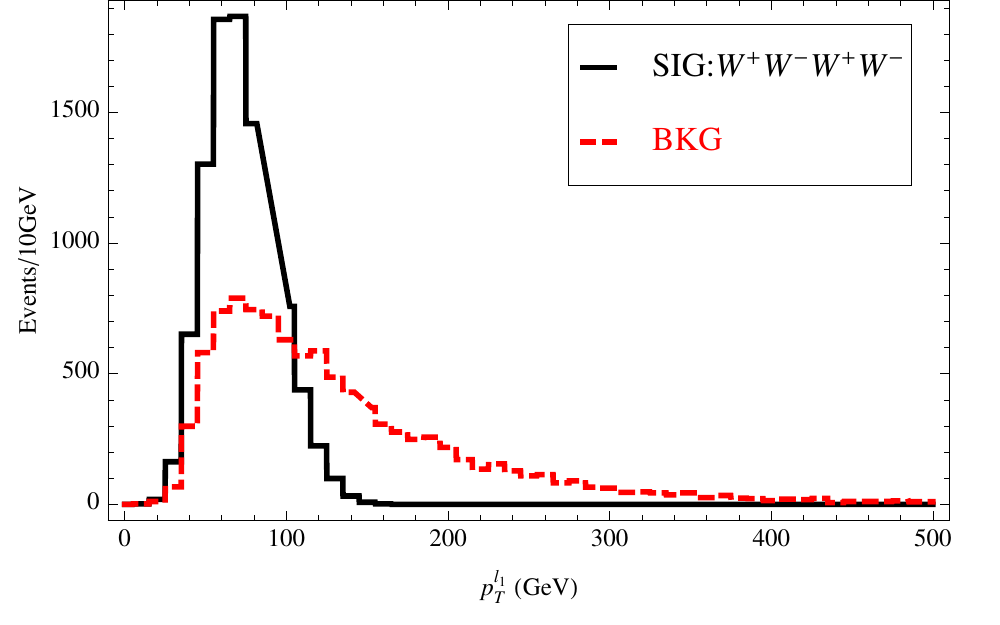}
\includegraphics[width=7.5cm]{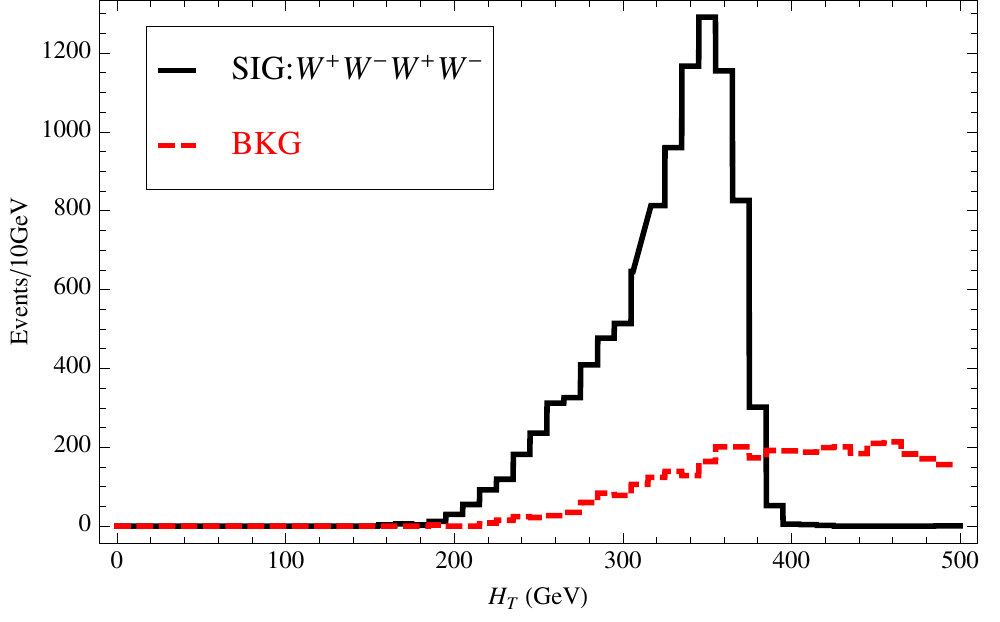}
\caption{The $p_T$ and $H_T$ distribution for signal $WW^{*} + WW^{*}$ and the corresponding background. The dashed red line is background and the solid black line is signal.}\label{fig:4W}
\end{figure}

In this section, we assume the signal process is $pp\rightarrow S\rightarrow HH\rightarrow(W^+l^-\nu)(Wjj)\rightarrow(l^-\nu l^+\nu)(jjjj)$. At the same time, the background contains: $pp\rightarrow (W^+l^-\nu)(Wjj)\rightarrow (l^-l^+\nu\nu)(jjjj)$. In order to choose the suitable cut conditions, we give the distribution of $p_T$ and $H_T$ in Fig.\ref{fig:4W}.

With the distribution shown in Fig.\ref{fig:4W}, we choose the cut conditions as following:
\begin{eqnarray}
\begin{split}
&p^{j_1}_T\leq 80GeV, p^{j_2}_T\leq 60GeV\\
&p^{l_1}_T\leq 120GeV, H_T\leq 400GeV
\end{split}
\end{eqnarray}

where the $H_T$ means the $p_T$ sum of all the leptons, jets and missing energy. To further suppress the background, we also impose the invariant mass conditions below,

\begin{eqnarray}
\begin{split}
&&m_{ll\nu\nu}, m_{4j} \in [115GeV,135GeV], m_{ll\nu\nu4j} \in [360GeV,440GeV].
\end{split}
\end{eqnarray}

After above cuts, the event number of signal and background are shown in Table.\ref{tab:4W}. The signal of this channel is very good, even before the cut the background cross-section is smaller than signal. The cut efficiency of the background vanishes after the cut, which means this channel is clean indeed. The event number of the signal is quite large that one expects very good statistics on the signal.

 \begin{table}[tbp]
\centering
\begin{tabular}{lcc}
\hline
event number &signal &BKG($l^-l^+\nu\nu jjjj$)\\ \hline
cross-section($pb$)&$4.4\times 10^{-2}$&0.0012\\
cut efficiency&  0.7632 &0\\
event number & $3.36 \times 10^4 $  & 0 \\
$\frac{S}{B}$ ratio&$\infty$\\ \hline
\end{tabular}
\caption{The cut flow table for of $WW^{*} + WW^{*}$ channel.}\label{tab:4W}
\end{table}

\subsection{Two Higgs decay to $\gamma\gamma+\gamma\gamma$}
\begin{figure}[ht]
\includegraphics[width=7.5cm]{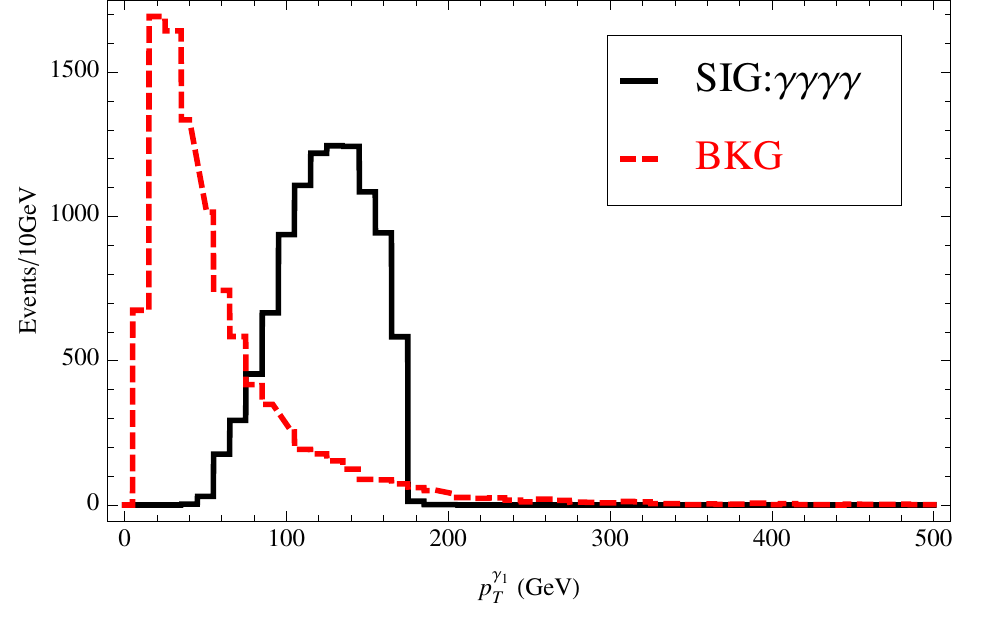}
\includegraphics[width=7.5cm]{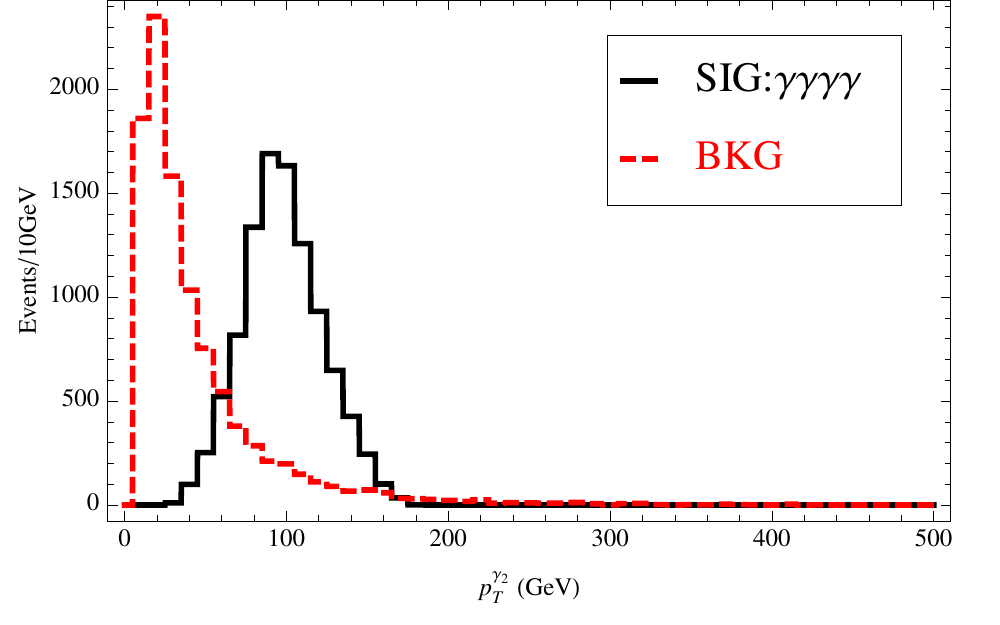}
\includegraphics[width=7.5cm]{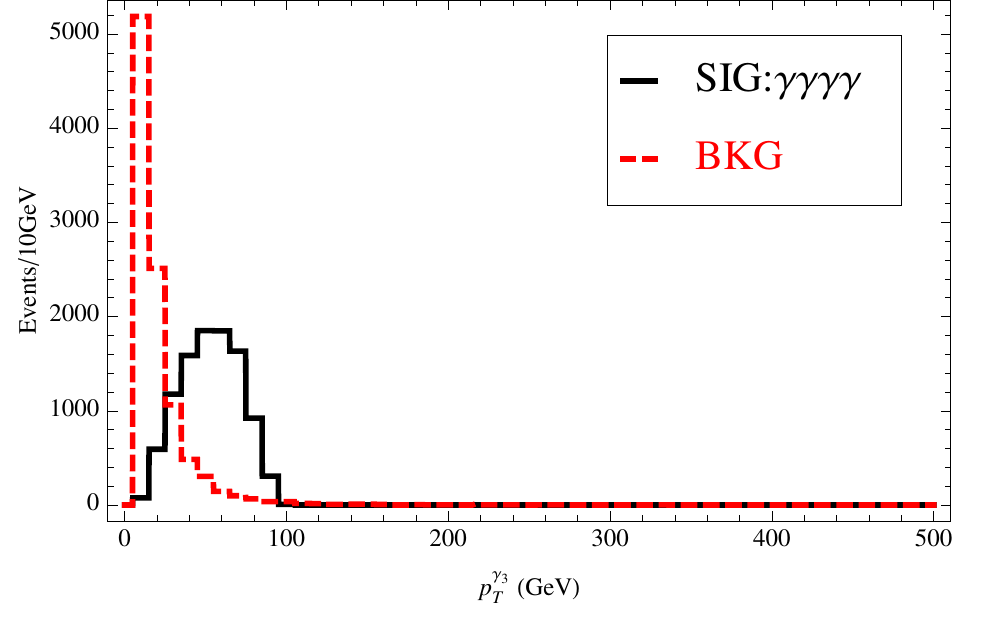}
\includegraphics[width=7.5cm]{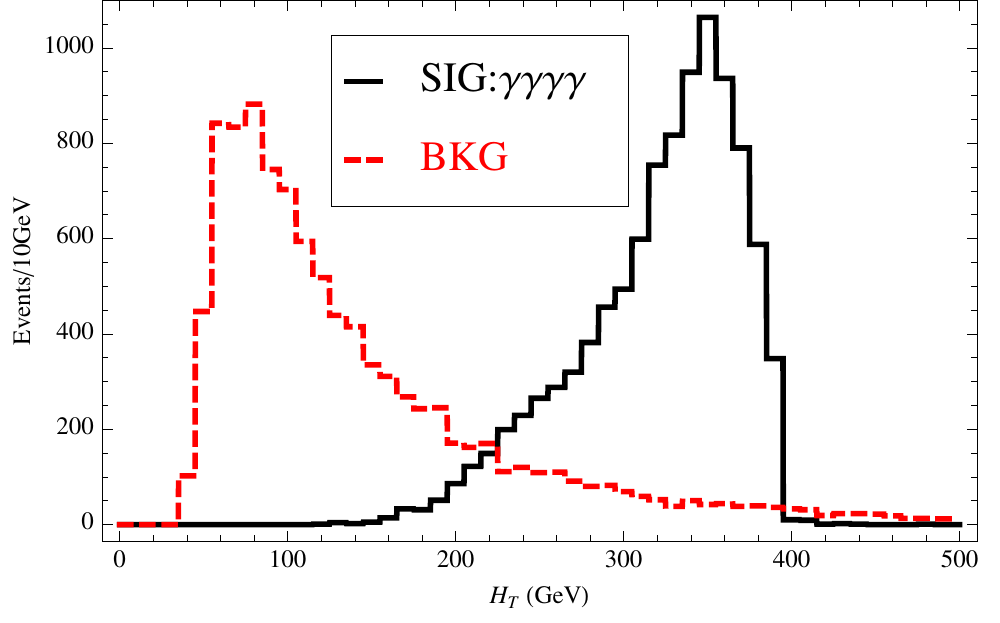}
\caption{The $p_T$ and $H_T$ distribution of signal and background for final state $\gamma\gamma+\gamma\gamma$. The dashed red line is background, while the solid black line is signal.}\label{fig:4a}
\end{figure}

In this section, we focus on the Higgs rare decay to photons. We assume the signal process is $pp\rightarrow S\rightarrow HH\rightarrow(\gamma\gamma)(\gamma\gamma)$. At the same time, the background contains prompt $\gamma$ production $pp\rightarrow \gamma\gamma\gamma\gamma$, and non-prompt gamma from jet faking which is hard to quantify at parton level. In order to choose the suitable cut conditions, we give the $p_T$ and $H_T$ distribution of photons in Fig.\ref{fig:4a}.

With the distribution shown in Fig.\ref{fig:4a}, we choose the cut conditions below,
\begin{eqnarray}
\begin{split}
&p^{\gamma_1}_T\geq 80GeV, p^{\gamma_2}_T\geq 60GeV\\
&p^{\gamma_3}_T\geq 30GeV, p^{\gamma_4}_T\geq 20GeV, H_T\geq 200GeV.
\end{split}
\end{eqnarray}

We also impose the invariant mass cut conditions as following, to suppress the background,

\begin{eqnarray}
\begin{split}
&&m_{\gamma\gamma} \in [115GeV,135GeV], m_{\gamma\gamma\gamma\gamma} \in [360GeV,440GeV].
\end{split}
\end{eqnarray}

After above cuts, we give the event number of signal and background in Table.\ref{tab:4a}. The cross-section for signal and background are at the similar order, while the cut efficiency is very different. The $S/B$ ratio of signal is good due to very small electroweak background. But the $\gamma\gamma\gamma\gamma$ signal also suffers the low statistics, same as in the $ZZ^{*} +ZZ^{*}$ channel. The non-prompt $\gamma$ from jet faking also makes this channel more difficult.

 \begin{table}[tbp]
\centering
\begin{tabular}{lcc}
\hline
event number &signal &BKG($\gamma\gamma\gamma\gamma$)\\ \hline
cross-section($pb$)&$5.29 \times 10^{-5}$&$2.0 \times 10^{-4}$\\
cut efficiency&  0.6683 &  0.0015\\
total event& 35 & 0.3 \\
$\frac{S}{B}$ ratio&117\\ \hline
\end{tabular}
\caption{The cut flow table for two Higgs decay to $\gamma\gamma\gamma\gamma$.}\label{tab:4a}
\end{table}

\subsection{Two Higgs decay to $\gamma\gamma +ZZ^{*}$}
\begin{figure}[ht]
\includegraphics[width=7.5cm]{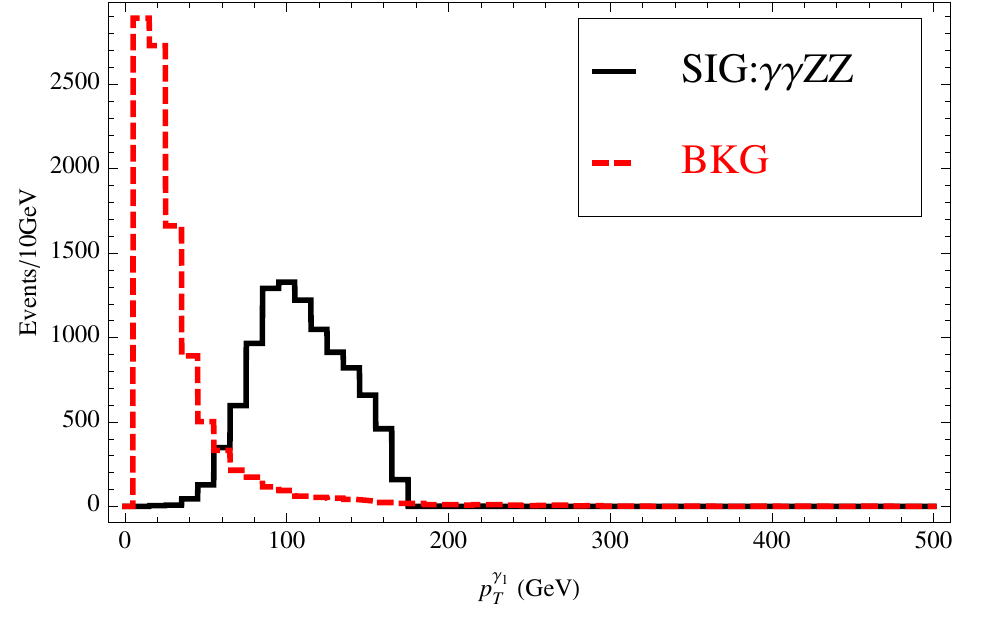}
\includegraphics[width=7.5cm]{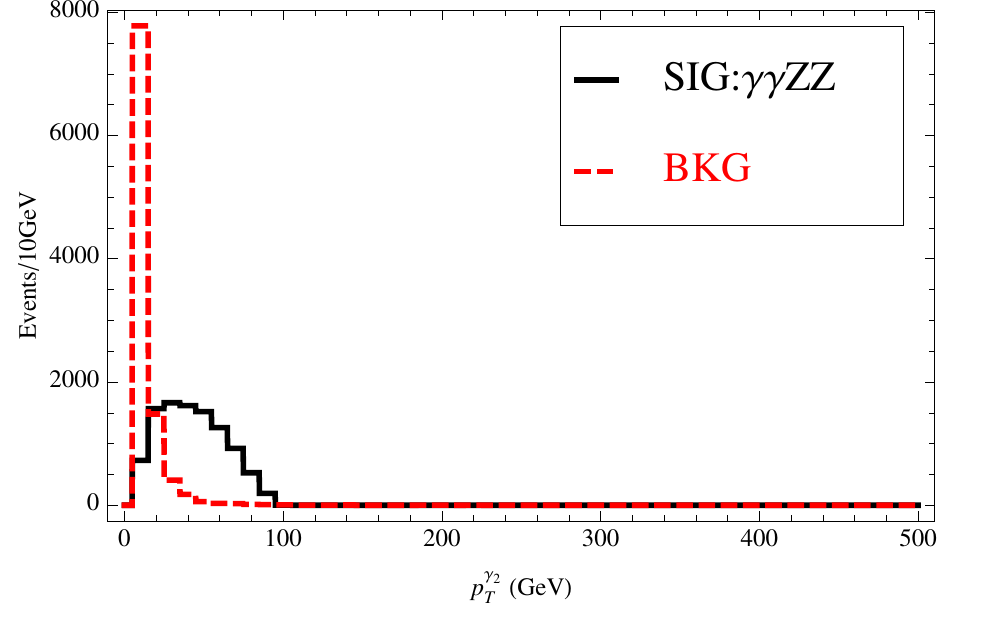}
\includegraphics[width=7.5cm]{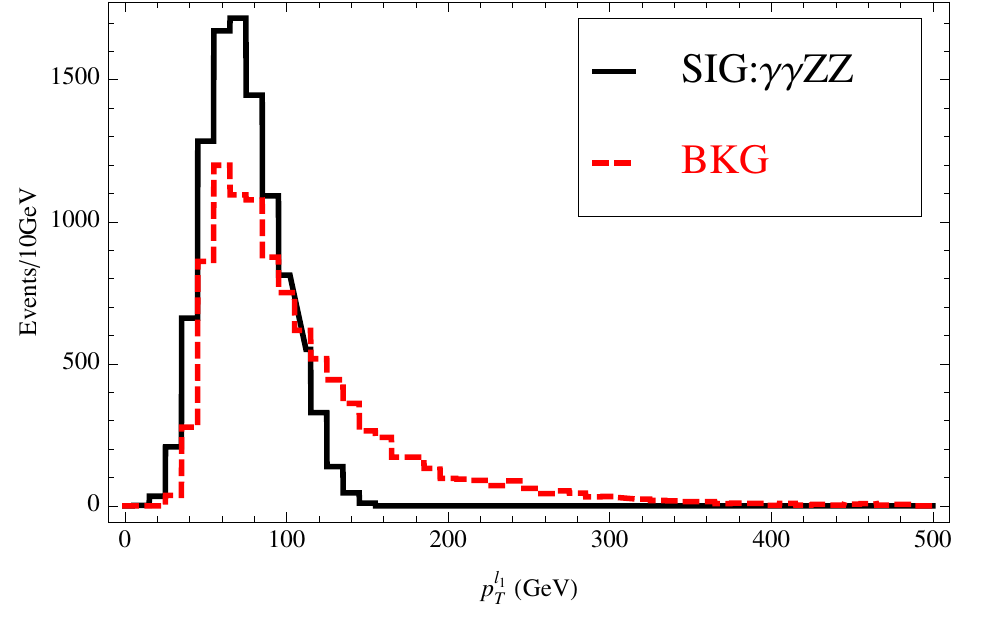}
\includegraphics[width=7.5cm]{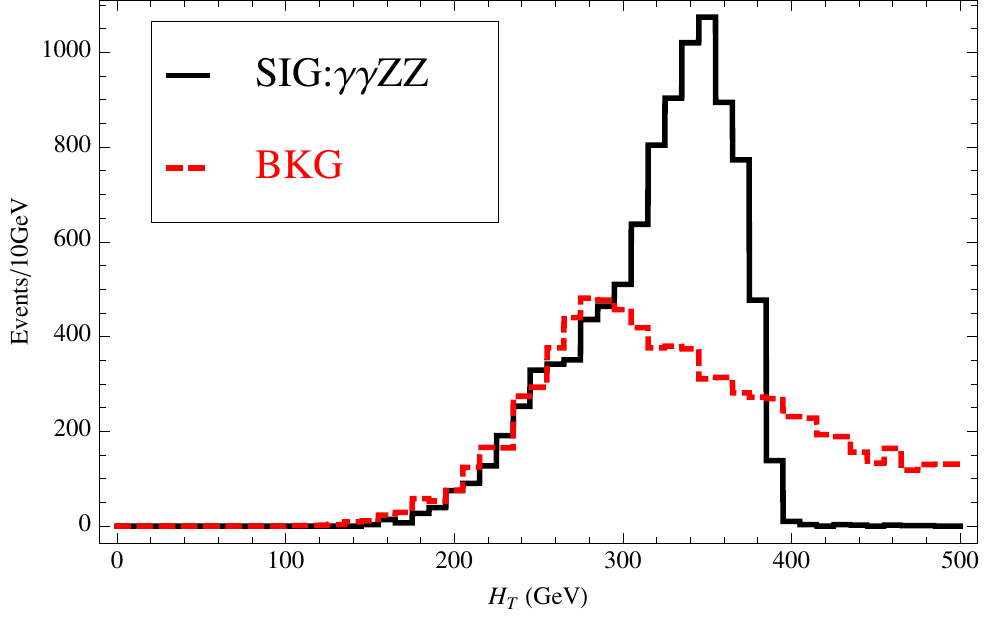}
\caption{The $p_T$ and $H_T$ distribution of gamma and leptons for signal and background from final state $\gamma\gamma l^-l^+l^-l^+$. The dashed red line is background, while the solid black line is signal.}\label{fig:2a2Z}
\end{figure}

In this section, we assume the signal process is $pp\rightarrow S\rightarrow HH\rightarrow(\gamma\gamma)(ZZ^{*})$ with Z decay leptonically. At the same time, the background contains: $pp\rightarrow \gamma\gamma l^-l^+l^-l^+$. In order to choose the suitable cut conditions, we give the $p_T$ and $H_T$ distribution of photons and leptons in the Fig.\ref{fig:2a2Z}.

With the distribution shown in Fig.\ref{fig:2a2Z}, we choose the cut conditions as following,
\begin{eqnarray}
\begin{split}
&p^{\gamma_1}_T\geq 60GeV, p^{\gamma_2}_T\geq 20GeV\\
&p^{l_1}_T\geq 20GeV, H_T\leq 400GeV,
\end{split}
\end{eqnarray}

where the $H_T$ means the $p_T$ sum of leptons and photons. We also propose the invariant mass cut conditions,

\begin{eqnarray}
\begin{split}
&&m_{\gamma\gamma} \in [115GeV,135GeV], m_{4l} \in [115GeV,135GeV] \\ &&m_{\gamma\gamma 4l} \in [360GeV,440GeV]
\end{split}
\end{eqnarray}

After above cuts, the event number of signal and background are shown in Table.\ref{tab:2a2Z}. The $\gamma\gamma +ZZ^{*}$ channel is very clean, that the electroweak background is very low even before the cut. But one should be cautious that the statistics for signal is quite low.

 \begin{table}[tbp]
\centering
\begin{tabular}{lcc}
\hline
event number &signal &BKG($\gamma\gamma l^-l^+l^-l^+$)\\ \hline
cross-section($pb$)&$1.23 \times 10^{-5}$& $2.495 \times 10^{-7}$\\
cut efficiency&0.7423 & $1\times10^{-4}$\\
event number & 9.1  & $2.459\times10^{-5}$ \\
$\frac{S}{B}$ ratio& $3.6\times10^5$\\ \hline
\end{tabular}
\caption{The cut flow table for two Higgs decay to $\gamma\gamma ZZ^{*}$ with Z boson decay to leptons. In the cross-section for background, we may have added prior $p_T$ requirement in the Madgraph to generate events more efficiently.}\label{tab:2a2Z}
\end{table}

\subsection{Two Higgs decay to $\gamma\gamma +WW^{*}$  }
\begin{figure}[ht]
\includegraphics[width=7.5cm]{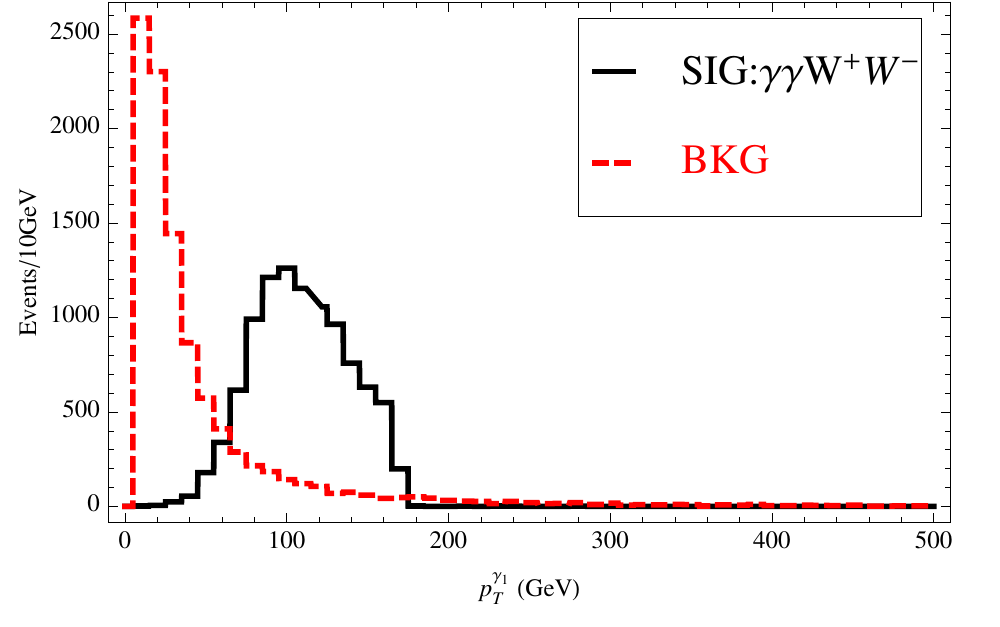}
\includegraphics[width=7.5cm]{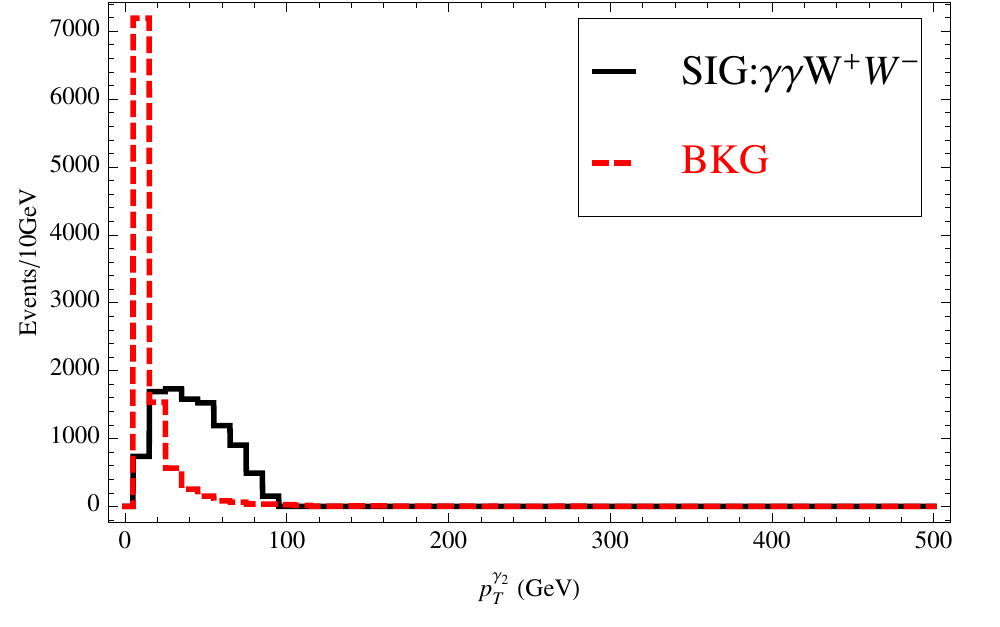}
\includegraphics[width=7.5cm]{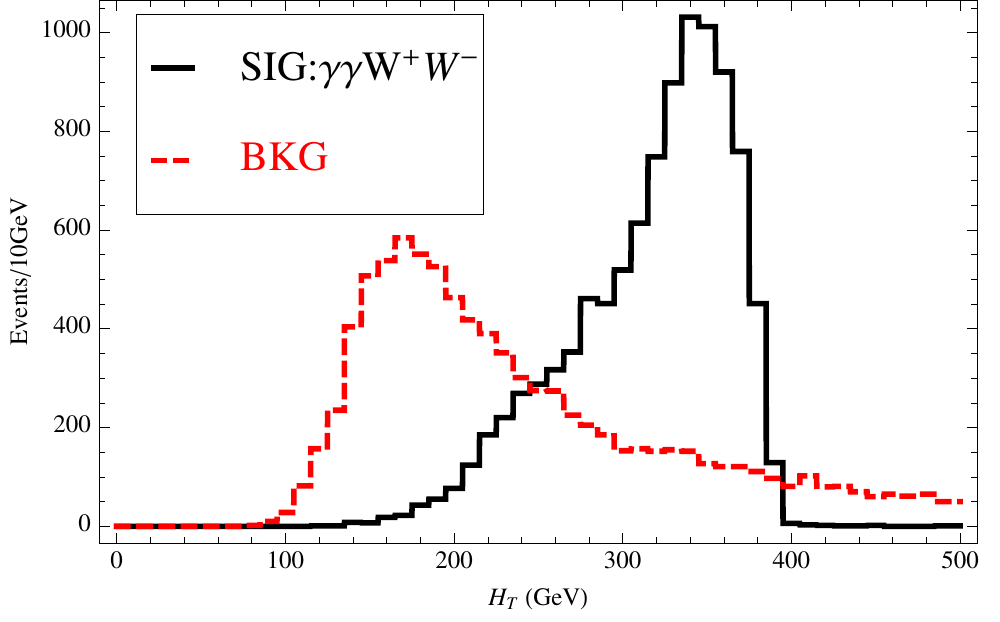}
\caption{The $p_T$ and $H_T$ distribution of signal and background for final state $\gamma\gamma +WW^{*}$ with W boson decay leptonically. The dashed red line is background , while the solid black line is signal.}\label{fig:2a2W}
\end{figure}

In this section, we assume the signal process is $pp\rightarrow S\rightarrow HH\rightarrow(\gamma\gamma)(WW^{*})$ with W boson decay leptonically. At the same time, the background contains $pp\rightarrow \gamma\gamma l^-l^+\nu\nu$. In order to choose the suitable cut conditions, we give the distribution of $p_T$ and $H_T$ for photons  and leptons in Fig.\ref{fig:2a2W}

With the distribution shown in Fig.\ref{fig:2a2W}, we choose the cut conditions as following,
\begin{eqnarray}
\begin{split}
&p^{\gamma_1}_T\geq 60GeV, p^{\gamma_2}_T\geq 20GeV\\
&p^{l_1}_T\geq 20GeV, H_T\geq 240GeV
\end{split}
\end{eqnarray}

where the $H_T$ means the $p_T$ sum of all the photons, leptons and missing energy. We also impose the invariant mass cut conditions,

\begin{eqnarray}
\begin{split}
&&m_{\gamma\gamma},m_{ll\nu\nu} \in [115GeV,135GeV], m_{\gamma\gamma ll\nu\nu} \in [360GeV,440GeV]
\end{split}
\end{eqnarray}

After above cuts, the event number of signal and background are shown in Table.\ref{tab:2a2W}. This channel has very good $S/B$ ratio and good statistics as well. The background is quite small even before the cuts, so this channel is quite clean.

 \begin{table}[tbp]
\centering
\begin{tabular}{lcc}
\hline
event number &signal &BKG($\gamma\gamma l^-l^+\nu\nu$)\\ \hline
cross-section($pb$)&$1.1 \times 10^{-3}$&$1.7 \times 10^{-4}$\\
cut efficiency& 0.796 &0\\
event number & 861 &  0 \\
$\frac{S}{B}$ ratio&$\infty$\\ \hline
\end{tabular}
\caption{The cut table for two Higgs decay to $\gamma\gamma WW^{*}$ with $W$ boson decay leptonically.}\label{tab:2a2W}
\end{table}

\subsection{Two Higgs decay to $ZZ^{*}+WW^{*}$}
\begin{figure}[!htb]
\centering
\includegraphics[width=7.5cm]{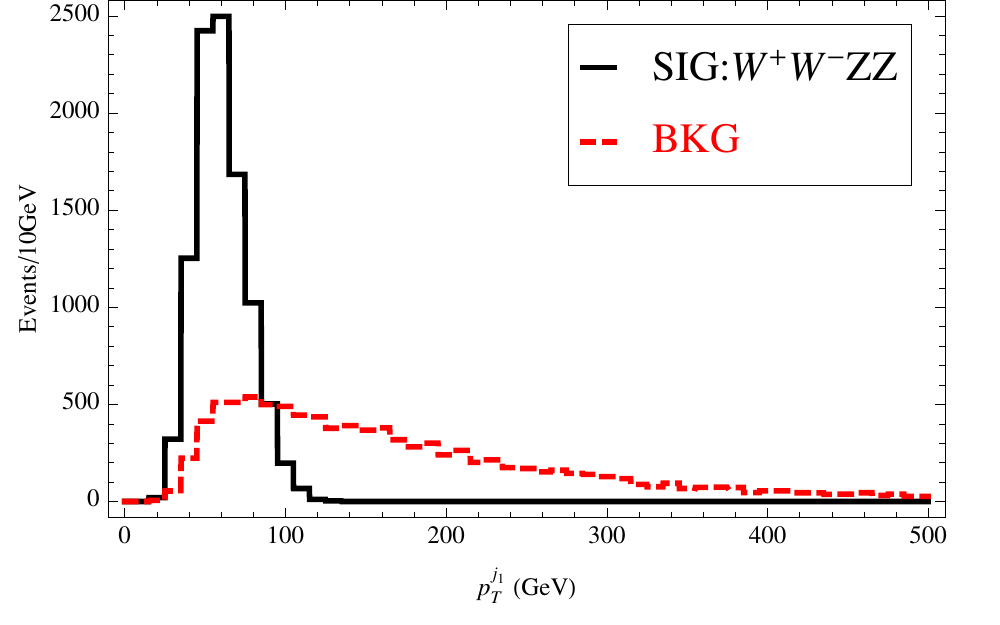}
\includegraphics[width=7.5cm]{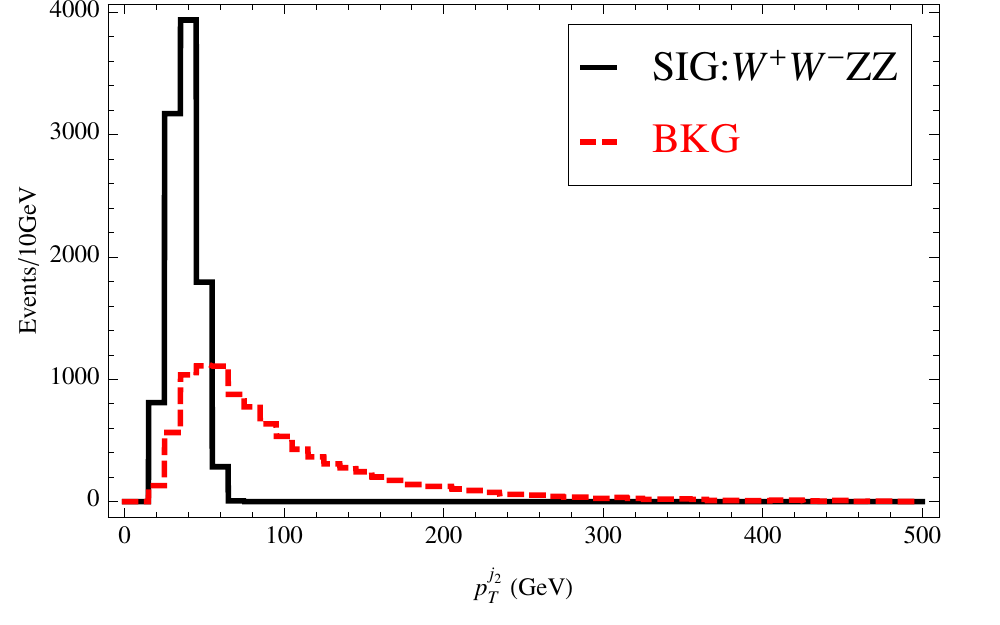}
\includegraphics[width=7.5cm]{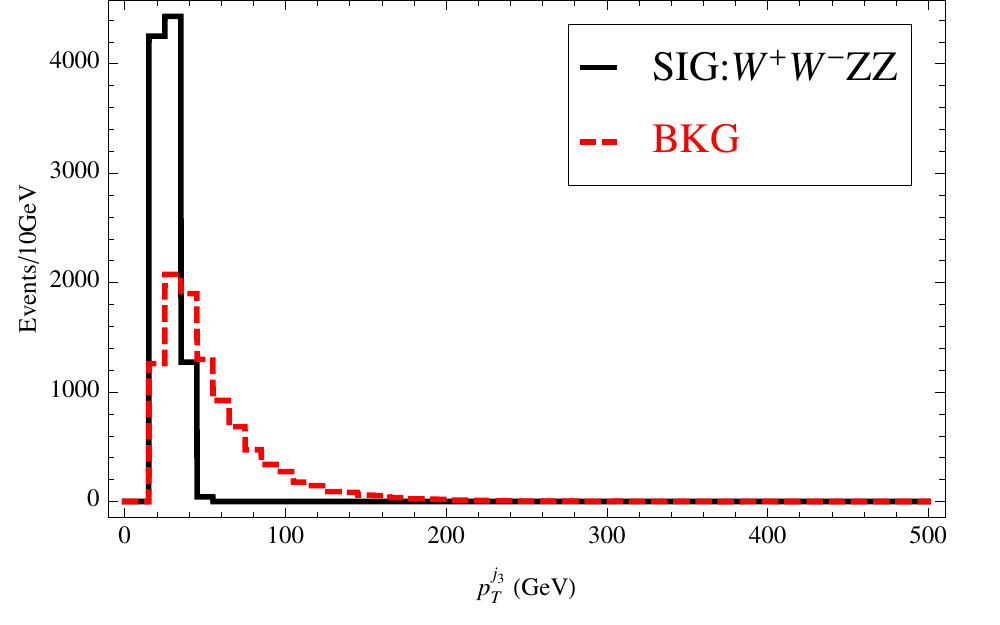}
\includegraphics[width=7.5cm]{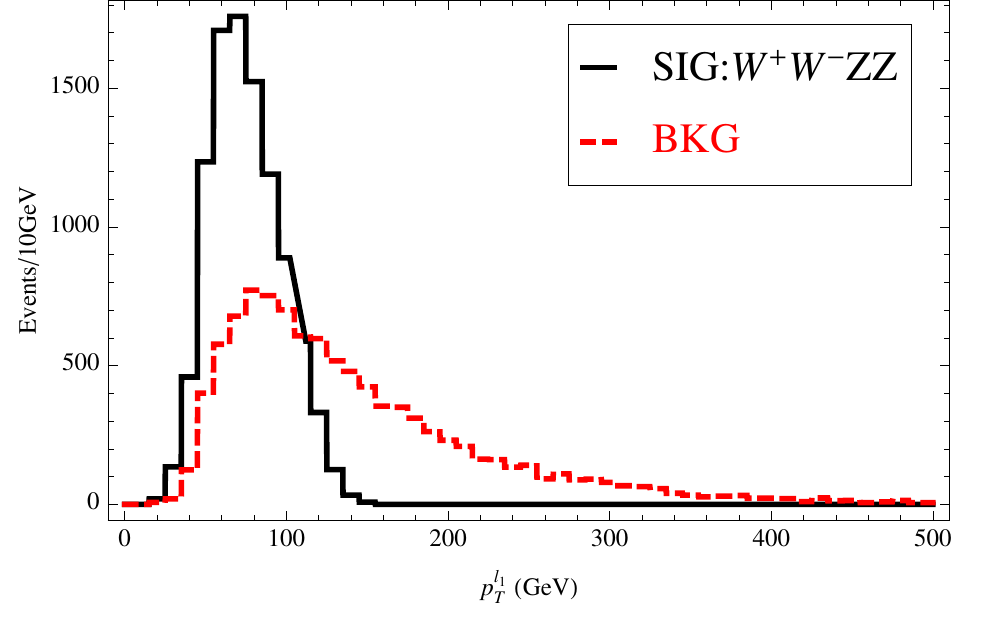}
\includegraphics[width=7.5cm]{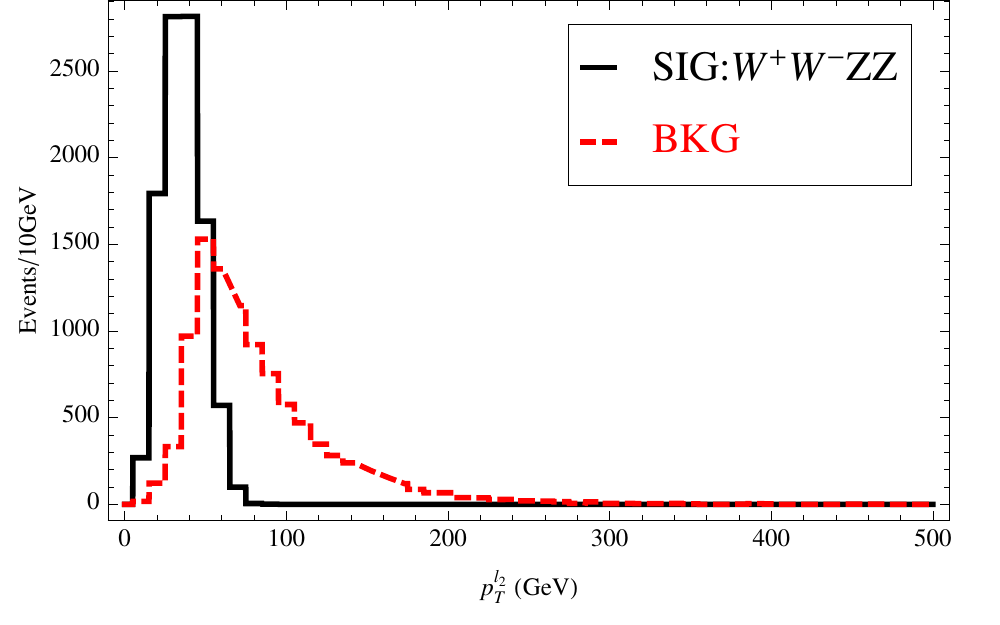}
\includegraphics[width=7.5cm]{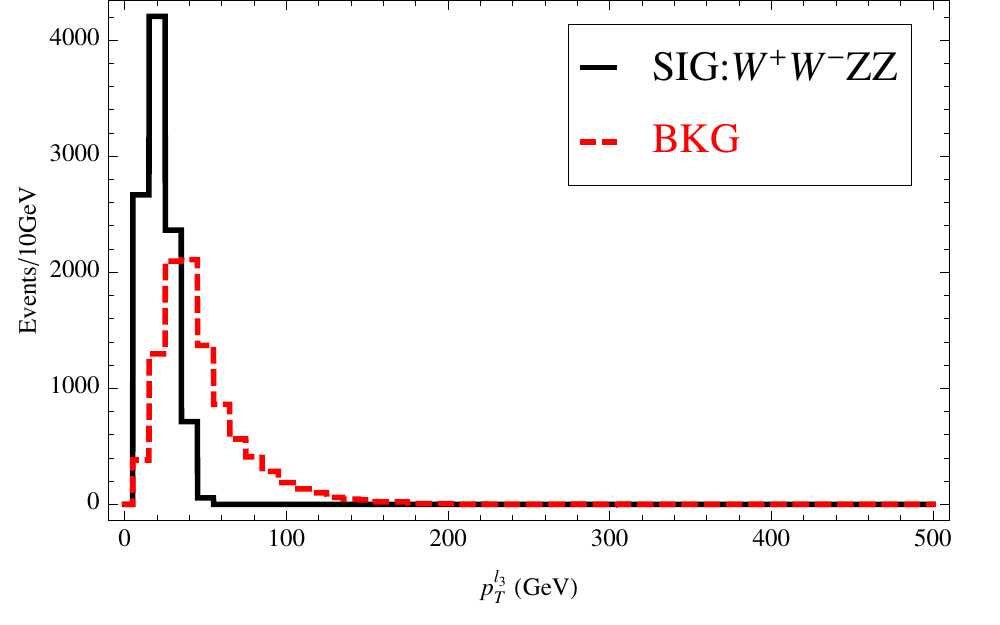}
\includegraphics[width=7.5cm]{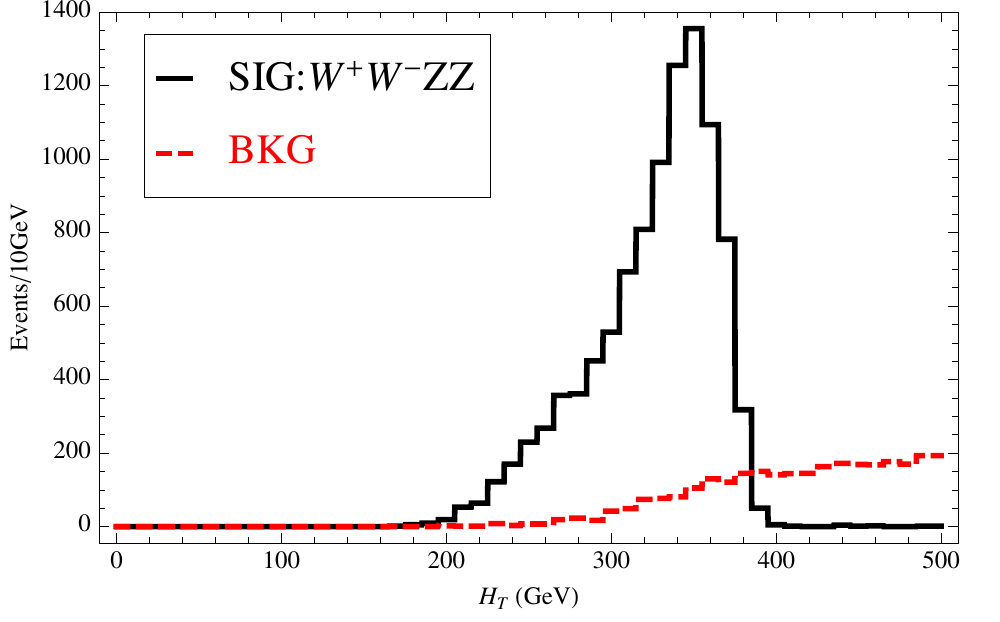}
\caption{The jets and leptons $p_T$ and $H_T$ distribution of signal and background for two Higgs final state $ZZ^{*}+WW^{*}$. The dashed red line is background, while the solid black line is signal.}
\label{fig:2Z2W}
\end{figure}

In this section, we assume the signal process is $pp\rightarrow S\rightarrow HH\rightarrow(ZZ^{*}WW^{*})$ with Z decay leptonically while W decay hadronically. At the same time, the background contains: $pp\rightarrow jjjjl^-l^+l^-l^+$. In order to choose the suitable cut conditions, we give the distribution of $p_T$ and $H_T$ in the Fig.\ref{fig:2Z2W}.

With the distribution shown in Fig.\ref{fig:2Z2W}, we choose the cut conditions as following:
\begin{eqnarray}
\begin{split}
&&p^{j_1}_T\leq 100GeV, p^{j_2}_T\leq 60GeV, p^{l_1}_T\leq 100GeV\\
&&p^{l_2}_T\leq 80GeV, p^{l_3}_T\leq 60GeV, H_T\leq 400GeV,
\end{split}
\end{eqnarray}

where the $H_T$ means the $p_T$ sum of jets and leptons. On the other hand, we also impose the invariant mass cut conditions,

\begin{eqnarray}
\begin{split}
&&m_{4j}, m_{4l}  \in [115GeV,135GeV], m_{4j4l} \in [360GeV,440GeV]
\end{split}
\end{eqnarray}

After above cuts, the event number of signal and background are shown in Table.\ref{tab:2Z2W}. The $S/B$ ratio of this channel is quite large because the small electroweak background. The signal has event number of 382, which is quite significant. One should note that we have not imposed the $Z$ veto cuts that only one pair of leptons has the invariant mass close to $Z$. If one imposed such cuts, the signal to background ratio could be higher, but at the cost of signal statistics. In our case, the signal is already good enough, that we suppress such cuts.

 \begin{table}[tbp]
\centering
\begin{tabular}{lcc}
\hline
event number &signal &BKG($jjjjl^-l^+l^-l^+$)\\ \hline
cross-section($pb$)&$5\times10^{-4}$&$3\times 10^{-5}$\\
cut efficiency &  0.76 &  $10^{-4}$  \\
event number &   382 &  $3\times 10^{-3} $\\
$\frac{S}{B}$ ratio&$1.27\times 10^{5}$\\ \hline
\end{tabular}
\caption{The cut table for two Higgs decay to $ZZ^{*} WW^{*}$ with $Z$ decay leptonically and $W$ decay hadronically.}\label{tab:2Z2W}
\end{table}

\subsection{Two Higgs decay to $\tau^{+}\tau^{-} \tau^{+}\tau^{-}$ }
\begin{figure}[ht]
\includegraphics[width=7.5cm]{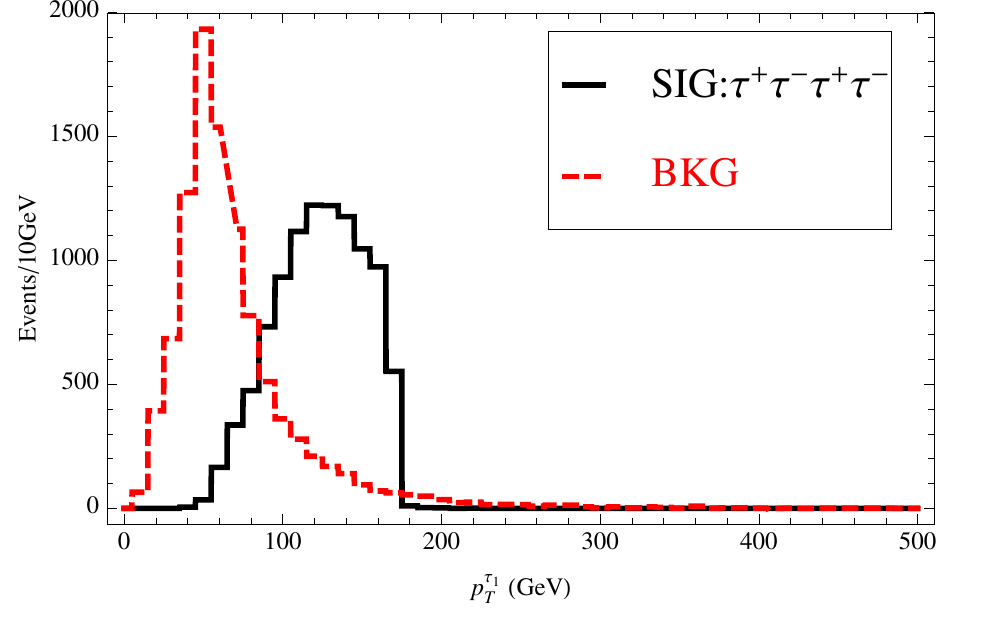}
\includegraphics[width=7.5cm]{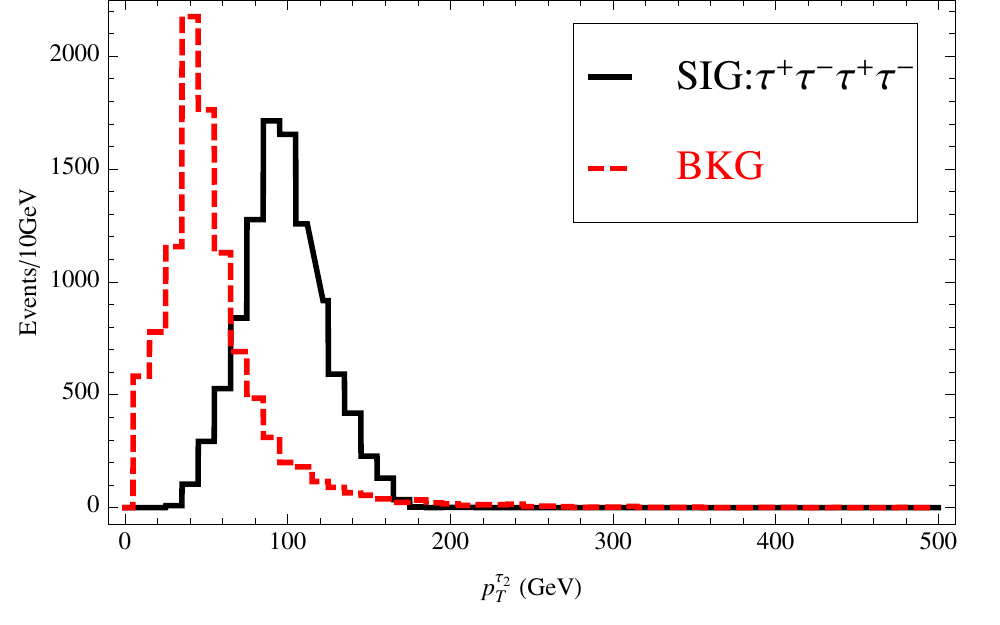}
\includegraphics[width=7.5cm]{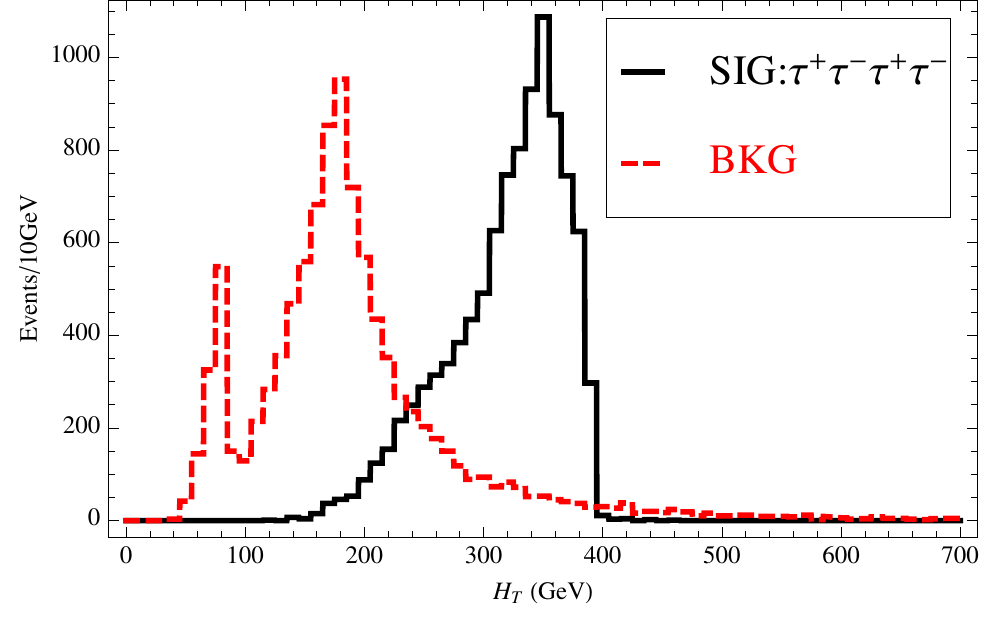}
\caption{The $p_T$ and $H_T$ distribution for signal $\tau^{+}\tau^{-} \tau^{+}\tau^{-}$ and corresponding background.  The dashed red line is background and the solid black line is signal.}\label{fig:4tau}
\end{figure}

In this section, we assume the signal process is $pp\rightarrow S\rightarrow HH\rightarrow(\tau^+\tau^-)(\tau^-\tau^+)$. At the same time, the background contains $pp\rightarrow \tau^-\tau^+\tau^-\tau^+$. In order to choose the suitable cut conditions, we give the distribution of $p_T$ and $H_T$ in the Fig.\ref{fig:4tau}.

With the distribution is shown is Fig.\ref{fig:4tau}, we choose the cut conditions as below:
\begin{eqnarray}
p^{\tau_1}_T\geq 80GeV, p^{\tau_2}_T\geq 60GeV, H_T\geq 240GeV
\end{eqnarray}

 where the $H_T$ is the $p_T$ sum for all the $\tau$ leptons. We also impose the invariant mass condition as following,

\begin{eqnarray}
\begin{split}
&&m_{\tau^-\tau^+} \in [115GeV,135GeV], m_{4 \tau} \in [360GeV,440GeV]
\end{split}
\end{eqnarray}

After above cuts, the event number for signal and background are shown in Table.\ref{tab:4tau}. One can see that the four $\tau$ signal has very good $S/B$ ratio and also large event number. In this channel, one should be careful about two things. First, we should note the multi-$\tau$ tagging efficiency in this signal, which would reduce both the signal and background. The hadronic $\tau$ tagging efficiency is about $50\%$ as in the Ref.\cite{Chatrchyan:2012zz}.  Second, in the jet backgrounds, there are possibility that they contribute to the non-prompt $\tau$ background. The misidentification rate is about $\sim 1\%$ as in the Ref.\cite{Chatrchyan:2012zz}. Comparing with the four jets background in the $4b$ channel, the non-prompt $4\tau$ contribution from jet background could be around $\sim 100$ events. Due to the above two factors, the $S/B$ ratio would not be as good as in the naive parton level estimation, of about $\frac{S}{B} \sim 100$ now but still a promising channel.

 \begin{table}[tbp]
\centering
\begin{tabular}{lccc}
event number &signal &BKG($\tau^-\tau^+\tau^-\tau^+$)\\ \hline
cross-section($pb$)&$4.1\times10^{-2}$&$7.153\times10^{-3}$\\
cut efficiency &  0.9077 & $5\times 10^{-4}$ \\
 $\tau_h$ tag   &      $(50\%)^{4}$      &  $(50\%)^{4}$ \\
event number &   2300 &  0.22\\
$\frac{S}{B}$ ratio&$1\times10^4$\\ \hline
\end{tabular}
\caption{The cut flow table for two Higgs decay to $\tau^{+}\tau^{-} \tau^{+}\tau^{-}$.}\label{tab:4tau}
\end{table}

\subsection{Two Higgs decay to $\tau^{+}\tau^{-}  b\bar{b}$}
\begin{figure}[ht]
\includegraphics[width=7.5cm]{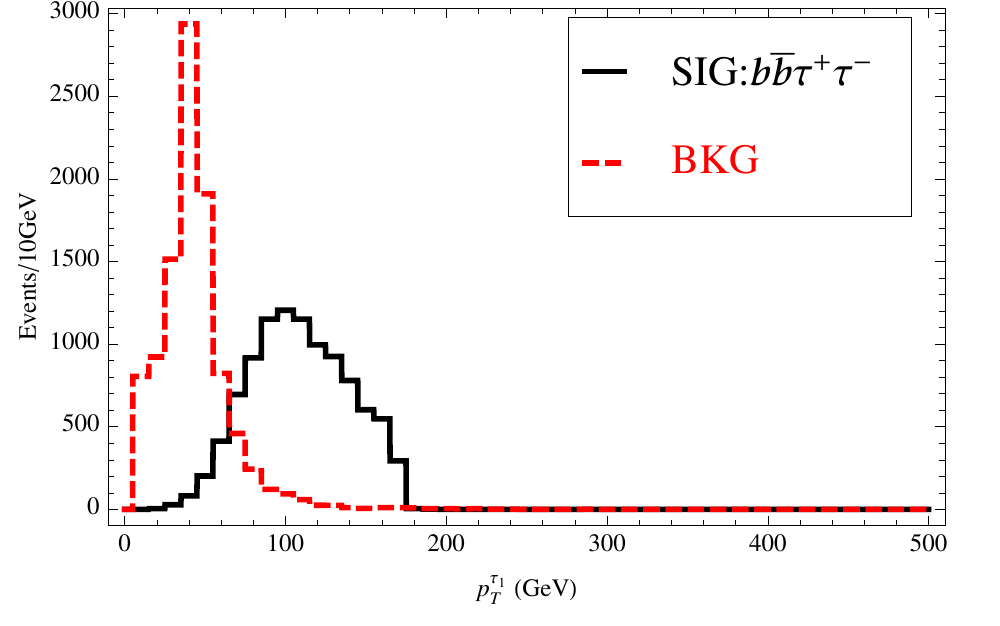}
\includegraphics[width=7.5cm]{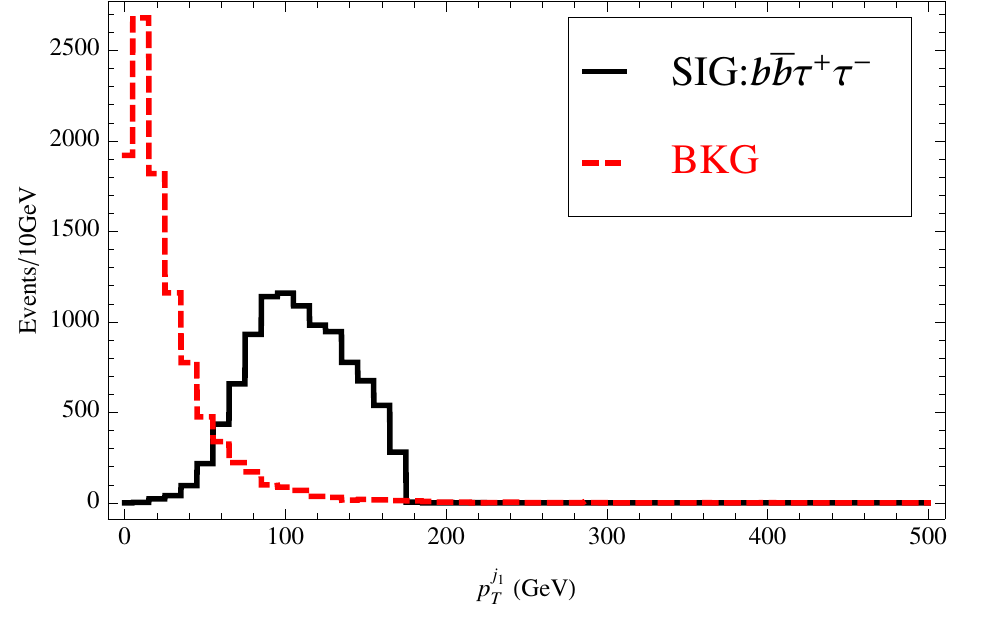}
\includegraphics[width=7.5cm]{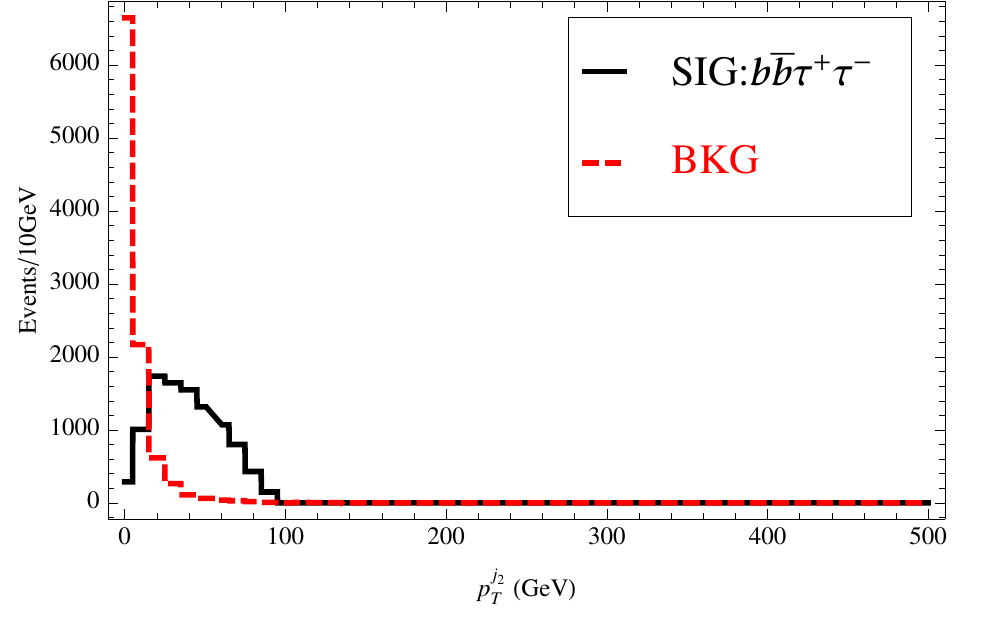}
\includegraphics[width=7.5cm]{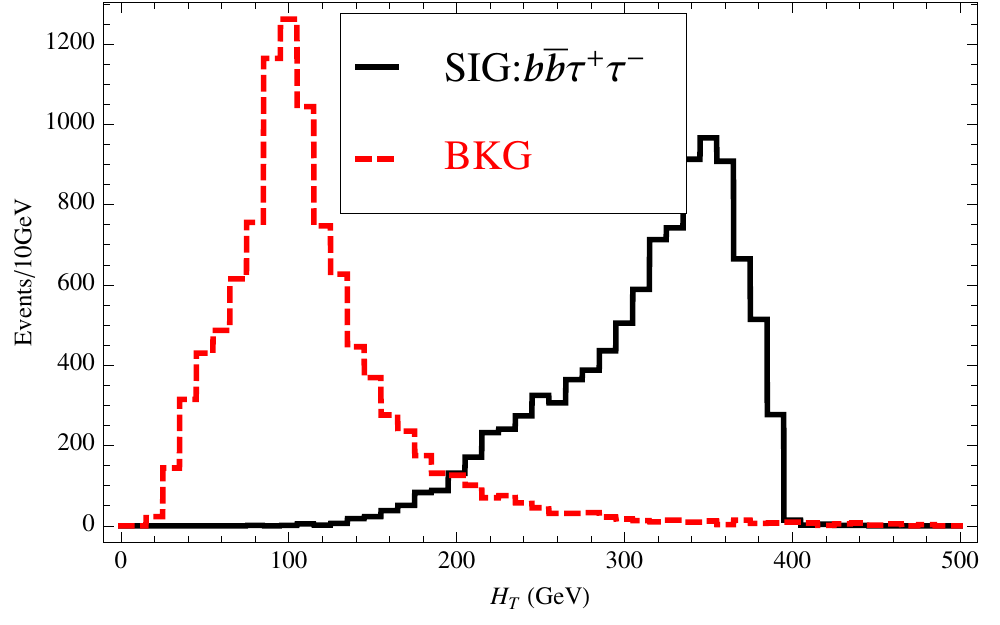}
\caption{The $p_T$ and $H_T$ distribution for signal $\tau^{+}\tau^{-}  b\bar{b}$ and corresponding background. The dashed red line is background and the solid black line is signal.}\label{fig:2b2tau}
\end{figure}

In this section, we assume the signal process is $pp\rightarrow S\rightarrow HH\rightarrow(\tau^-\tau^+)(b\bar{b})$. At the same time, the background contains jets plus two tau and top pair where top decay to tau \footnote{In the Ref.\cite{No:2013wsa}, the authors have also explored the same final states for a resonant heavy Higgs, where they find the $S/B \sim 1$.  The main difference is our signal strength is 25 times larger than them from the cross-section. The S particle has large coupling to the gluon which may benefit from new color particle in the loop, while their heavy Higgs coupling to gluon only comes from the mixing to SM Higgs. The other difference may come from the difference between the parton simulation and PGS simulation, where we benefit from no jet fragmentation, hadronization and energy smearing in the invariant mass cuts. We have got similar ratio between top pair background and jets plus two tau background comparing with their results.}. In order to choose the suitable cut conditions, we give the distribution of $p_T$ and $H_T$ in the Fig.\ref{fig:2b2tau}.

With the distribution is shown is Fig.\ref{fig:2b2tau}, we choose the cut conditions as following,
\begin{eqnarray}
\begin{split}
&&p^{j_1}_T\geq 60GeV, p^{j_2}_T\geq 20GeV, p^{\tau_1}_T\geq 40GeV\\
&&p^{\tau_2}_T\geq 20GeV, H_T\geq 200GeV
\end{split}
\end{eqnarray}

where the $H_T$ means the $p_T$ sum of jets and hadronic taus. For top pair background, we add an extra cuts that $MET < 50$GeV. We also impose the invariant mass cut which is crucial to cut down the background.

\begin{eqnarray}
\begin{split}
&&m_{jj}, m_{\tau^-\tau^+} \in [115GeV,135GeV], m_{2j2\tau} \in [360GeV,440GeV]
\end{split}
\end{eqnarray}

After above cuts, the event number for signal and background are shown in Table.\ref{tab:2b2tau}. This time the $S/B$ ratio is about two hundreds which is quite good. We also need to consider the misidentification of the hadronic $\tau$ from the jets background. Same as in the previous section, we can estimate this background from $4b$ channel that the jet backgrounds would contribute about $\sim 100$ events which is still much smaller than signal. In this case, the $S/B$ ratio will go down to about $\sim 100$, which is still a good channel to look at.

 \begin{table}[tbp]
\centering
\begin{tabular}{lcccc}
\hline
event number &signal &BKG($bb\tau^-\tau^+$)&BKG($jj\tau^- \tau^+$)& BKG($t\bar{t}$) \\\hline
cross-section($pb$)&0.736 & 2.59  &  21.61 & 4.53\\
cut efficiency &  0.7631 &  $1 \times 10^{-4}$ & $7 \times 10^{-4}$ & $5\times10^{-4}$\\
b-tag&0.36&0.36&$4\times10^{-4}$& 0.36\\
 $\tau_h$ tag   &      $(50\%)^{2}$      &  $(50\%)^{2}$  &  $(50\%)^{2}$ & $(50\%)^{2}$\\
event number &   $5.1 \times 10^{4}$ & 15 & 1.5 & 204\\
$\frac{S}{B}$ ratio& 231 \\ \hline
\end{tabular}
\caption{The cut flow table for two Higgs decay to $b\bar{b} \tau^-\tau^+$. In the cross-section for background, we may have added prior $p_T$ requirement in the Madgraph to generate events more efficiently. }\label{tab:2b2tau}
\end{table}

\subsection{Two Higgs decay to $\tau^-\tau^+  \gamma\gamma$}
\begin{figure}[ht]
\includegraphics[width=7.5cm]{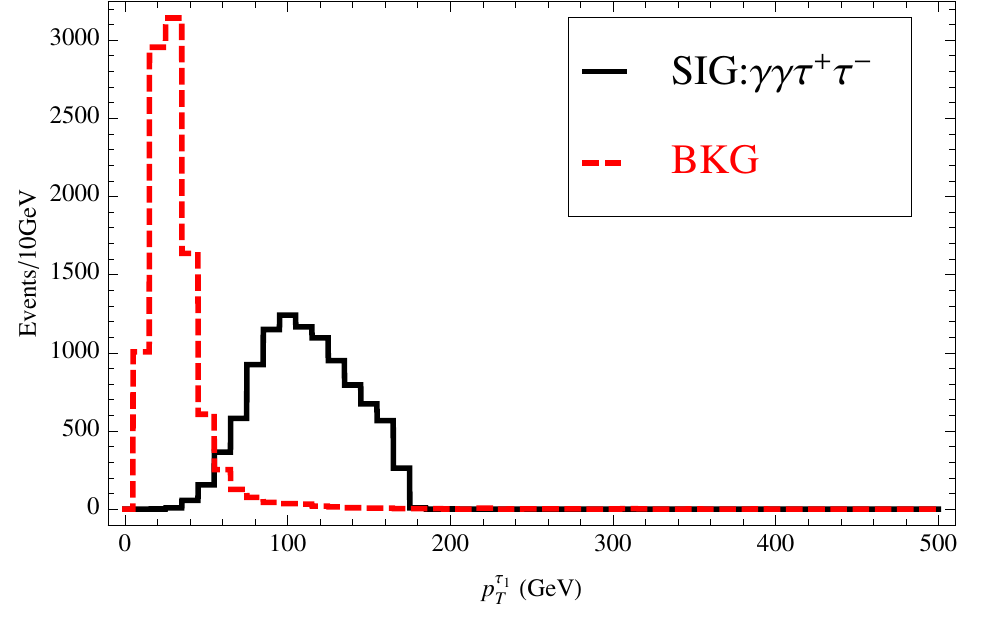}
\includegraphics[width=7.5cm]{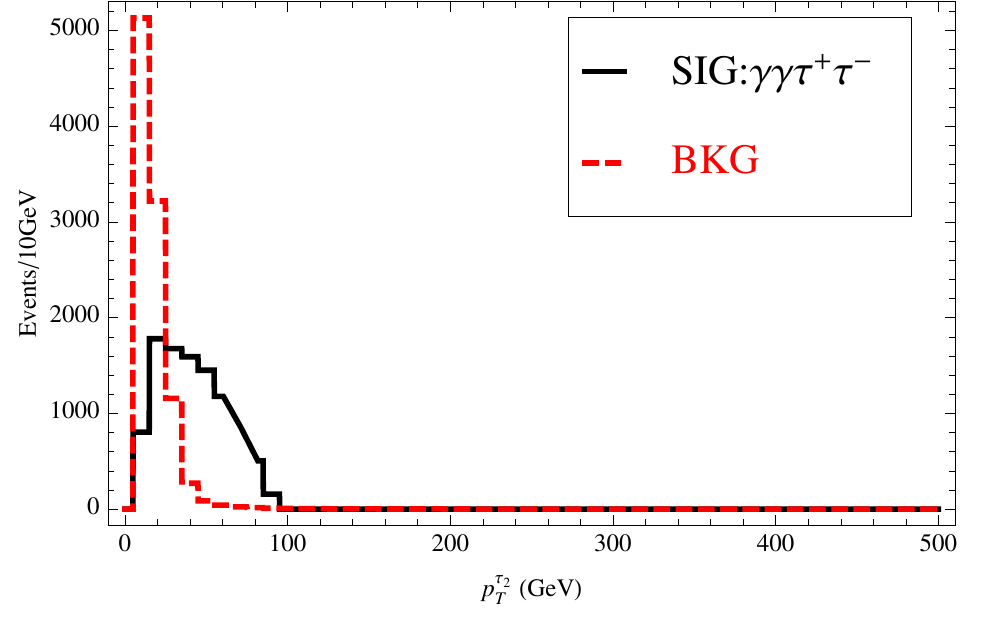}
\includegraphics[width=7.5cm]{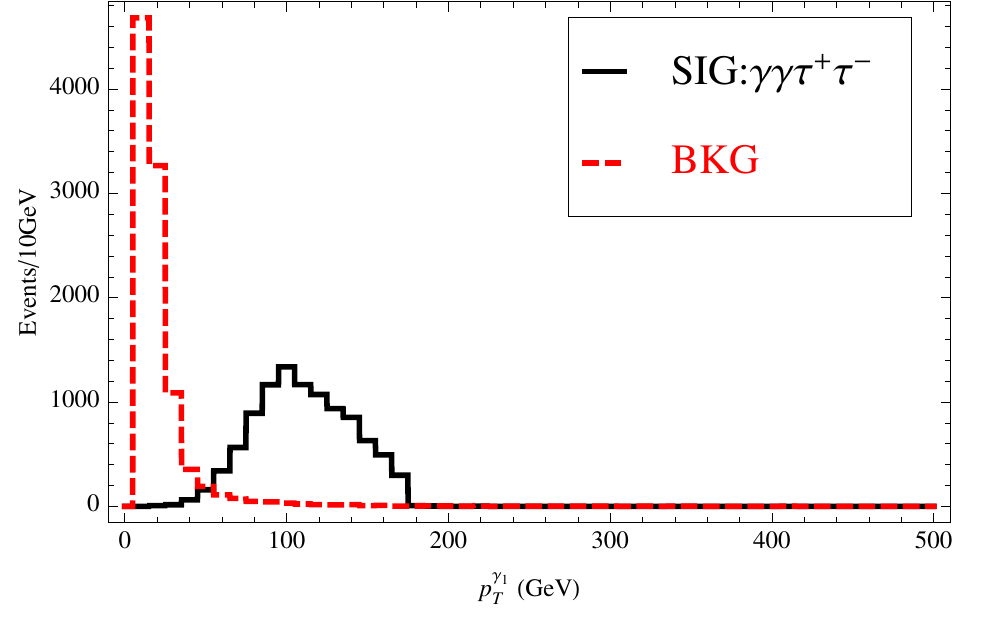}
\includegraphics[width=7.5cm]{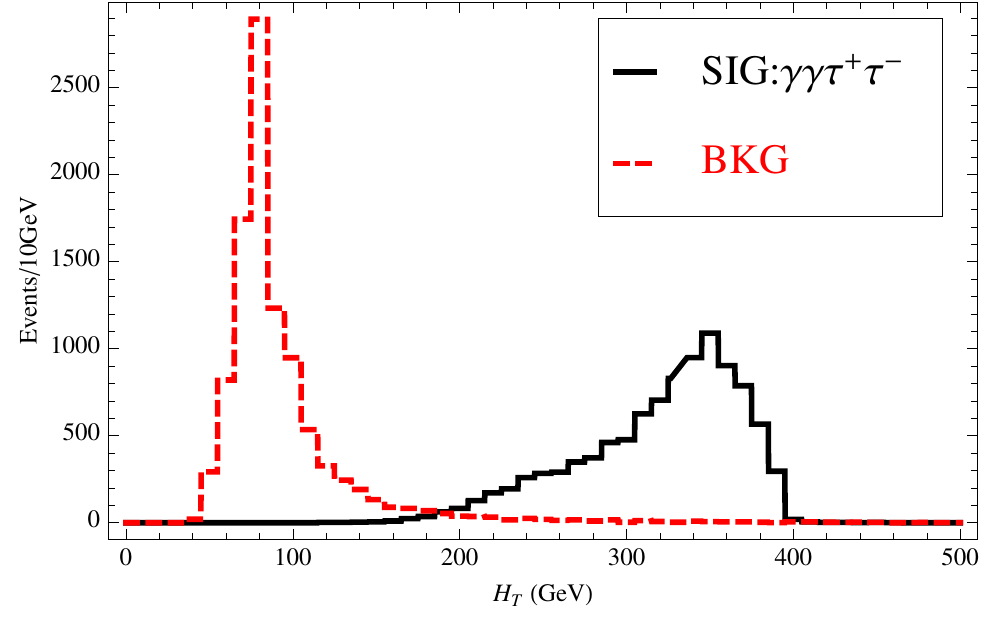}
\caption{The $p_T$ and $H_T$ distribution for signal $\tau^-\tau^+  \gamma\gamma$ and the corresponding background. The dashed red line is background and the solid black line is signal.}\label{fig:2tau2a}
\end{figure}

In this section, we assume the signal process is $pp\rightarrow S\rightarrow HH\rightarrow(\tau^+\tau^-)(\gamma\gamma)$. At the same time, the background contains  $pp\rightarrow \tau^-\tau^+\gamma\gamma$. In order to choose the suitable cut conditions, we give the distribution of $p_T$ and $H_T$ in Fig.\ref{fig:2tau2a}

With the distribution is shown in Fig.\ref{fig:2tau2a}, we choose the cut conditions as following,
\begin{eqnarray}
p^{\tau_1}_T\geq 60GeV, p^{\tau_2}_T\geq 20GeV, p^{\gamma_1}_T\geq 40GeV,H_T\geq 200GeV
\end{eqnarray}

where the $H_T$ means the $p_T$ sum of $\tau$ leptons and photons. We also impose the invariant mass cut conditions to further suppress the background,

\begin{eqnarray}
\begin{split}
&&m_{\tau^-\tau^+}, m_{\gamma\gamma} \in [115GeV,135GeV], m_{\tau^-\tau^+\gamma\gamma} \in [360GeV,440GeV]
\end{split}
\end{eqnarray}

After the above cuts, the event number for signal and background are shown in Table.\ref{tab:2tau2a}. In this channel, the electroweak background is quite small, with a very small cut efficiency. The background cross-section before the cut is also small, so the prompt background can be safely neglected. However, one should consider the non-prompt $\tau$ from jet misidentification. We estimate such background from the $b\bar{b} \gamma\gamma$ channel, substituting b-tag efficiency by the $\tau$ misidentification tagging efficiency, and since the kinematic cuts are similar, we can estimate that such non-prompt background can have about $\sim 10$ events which is still much smaller than signal. After counting the non-prompt background, we estimate that the  $S/B$ ratio could be around $\sim 60$ which is not bad.

 \begin{table}[tbp]
\centering
\begin{tabular}{lcc}
\hline
event number &signal &BKG($\tau^-\tau^+\gamma\gamma$)\\ \hline
cross-section($pb$)&$2.92 \times10^{-3}$& $9.688\times10^{-4}$\\
cut efficiency &  0.8774 & $4\times10^{-4}$ \\
$\tau_h$ tag   &      $(50\%)^{2}$      &  $(50\%)^{2}$   \\
event number &  640  &  0.1\\
$\frac{S}{B}$ ratio& $6.4\times10^{3}$\\ \hline
\end{tabular}
\caption{The cut flow table for two Higgs decay to $\tau^-\tau^+ \gamma\gamma$. In the cross-section for background, we may have added prior $p_T$ requirement in the Madgraph to generate events more efficiently.}\label{tab:2tau2a}
\end{table}

\subsection{Two Higgs decay to $\tau^-\tau^+ W^-W^+$}
\begin{figure}[ht]
\includegraphics[width=7.5cm]{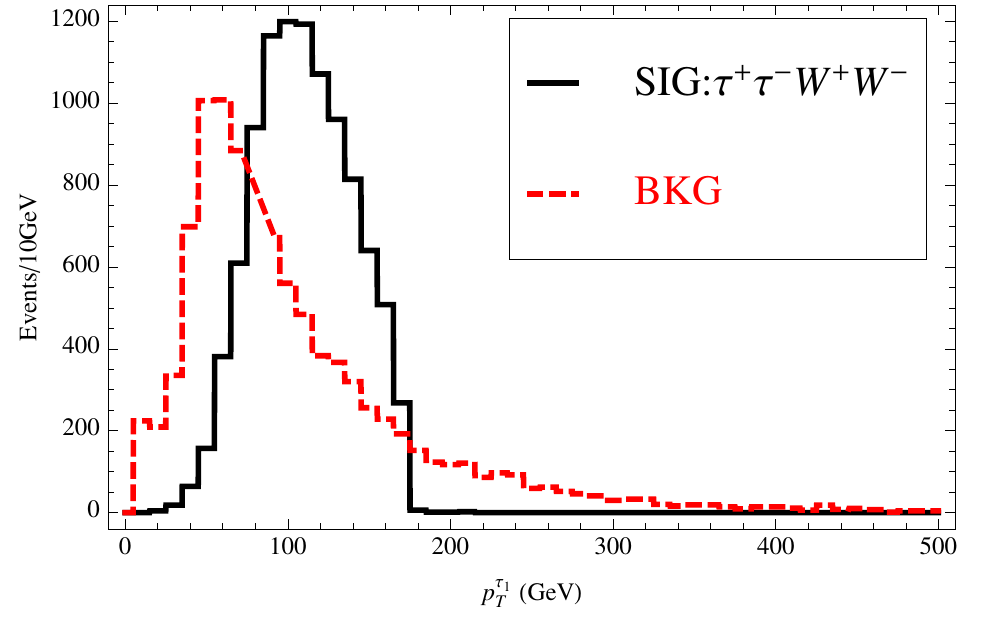}
\includegraphics[width=7.5cm]{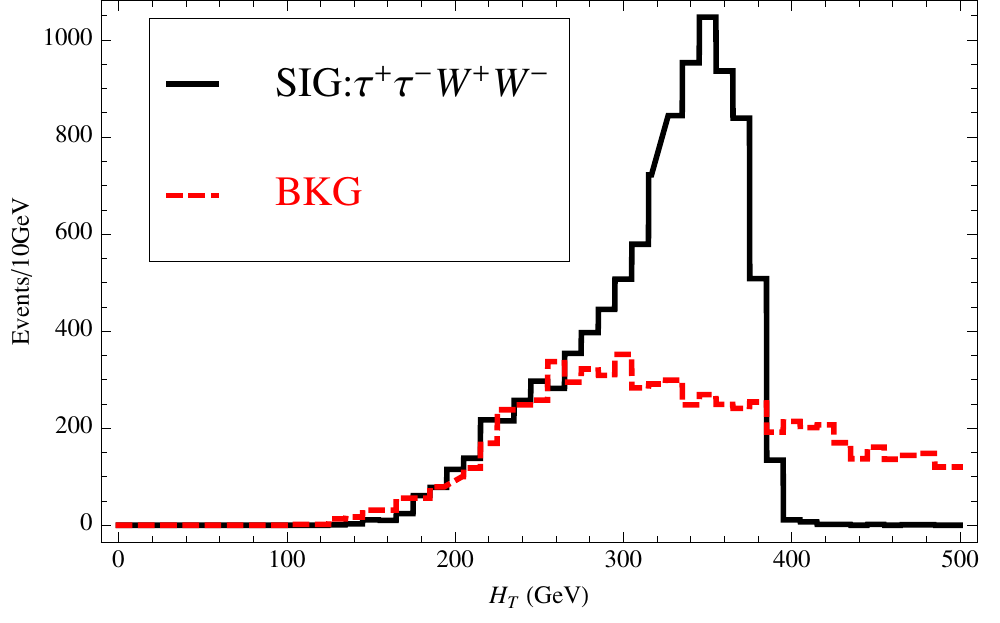}
\caption{The $p_T$ and $H_T$ distribution for signal $\tau^-\tau^+ W^-W^+$ and the corresponding background. The dashed red line is background and the solid black line is signal.}\label{fig:2tau2W}
\end{figure}

In this section, we assume the signal process is $pp\rightarrow S\rightarrow HH\rightarrow(\tau^-\tau^+)(W^{\pm}l^{\mp}\nu)$ $\rightarrow(\tau^-\tau^+)(l^-\nu l^+\nu)$. At the same time, the background contains $pp\rightarrow \tau^-\tau^+(W^{\pm}l^{\mp}\nu)$ $\rightarrow(\tau^-\tau^+)(l^-\nu l^+\nu)$. In order to choose the suitable cut conditions, we give the distribution of $p_T$ and $H_T$ in Fig.\ref{fig:2tau2W}.

With the distribution shown in Fig.\ref{fig:2tau2W}, we choose the cut conditions as following,
\begin{eqnarray}
\begin{split}
&&p^{\tau_1}_T\geq 80GeV, H_T\leq 400GeV
\end{split}
\end{eqnarray}

We also impose the invariant mass cut conditions to further increase the $S/B$ ratio.

\begin{eqnarray}
\begin{split}
&&m_{WW}, m_{\tau^-\tau^+}\in [115GeV,135GeV], m_{2W2\tau} \in [360GeV,440GeV]
\end{split}
\end{eqnarray}

After above cuts, the event number for signal and background are shown in Table.\ref{tab:2tau2W}. One might worry about the misidentification of $\tau$ as in the previous section. We look back into the $b\bar{b}  W^-W^+$ channel, after apply the misidentification $\tau$ tagging efficiency, the background cross-section for jets plus $l^-l^+\nu\nu$ is already much smaller than the signal. With a very small cut efficiency it have, we could be quite sure the $S/B$ ratio will be very good. One thing should be mention that in the measurement the missing energy should be reconstructed by the on-shell mass conditions, which could reduce the signal efficiency and also bring more background events. But still this is a very promising channel to see.

 \begin{table}[tbp]
\centering
\begin{tabular}{lcc}
\hline
event number &signal &BKG($l^-l^+\nu\nu\tau^-\tau^+$)\\ \hline
cross-section($pb$)& 0.03 & $1.078\times10^{-5}$\\
cut efficiency &  0.8740 &  $2.3\times10^{-4}$  \\
$\tau_h$ tag   &      $(50\%)^{2}$      &  $(50\%)^{2}$   \\
event number &  6567  &  $6\times10^{-4}$ \\
$\frac{S}{B}$ ratio& $1.09\times10^{8}$\\ \hline
\end{tabular}
\caption{The cut flow table for two Higgs decay to $\tau^-\tau^+ W^-W^+$. Note the leptonic decay BR for $W$ has been included in the signal cross-section. In the cross-section for background, we may have added prior $p_T$ requirement in the Madgraph to generate events more efficiently.}\label{tab:2tau2W}
\end{table}

\subsection{Two Higgs decay to $\tau^-\tau^+ ZZ$}
\begin{figure}[ht]
\includegraphics[width=7.5cm]{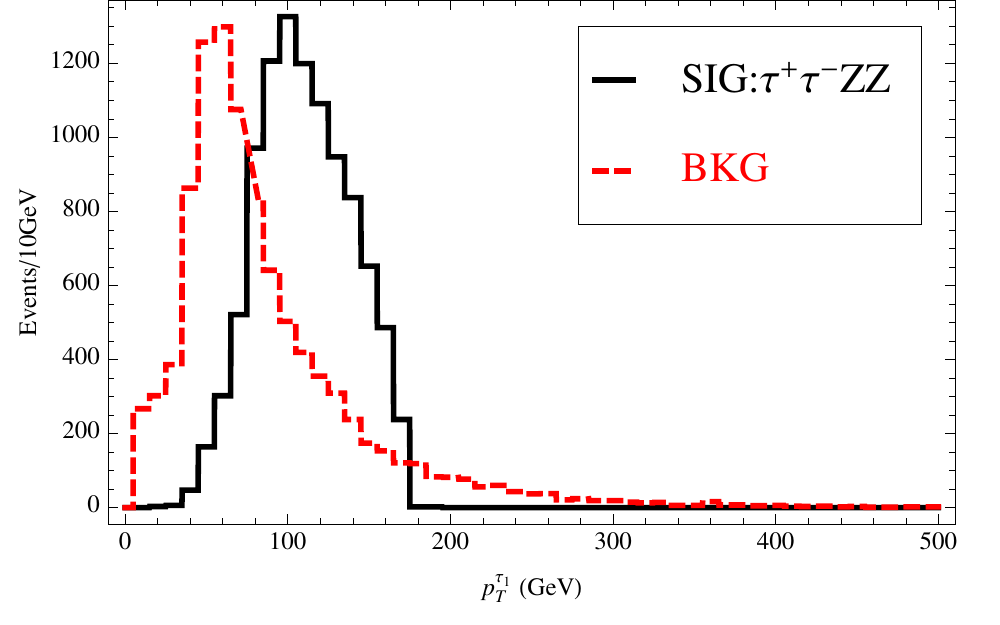}
\includegraphics[width=7.5cm]{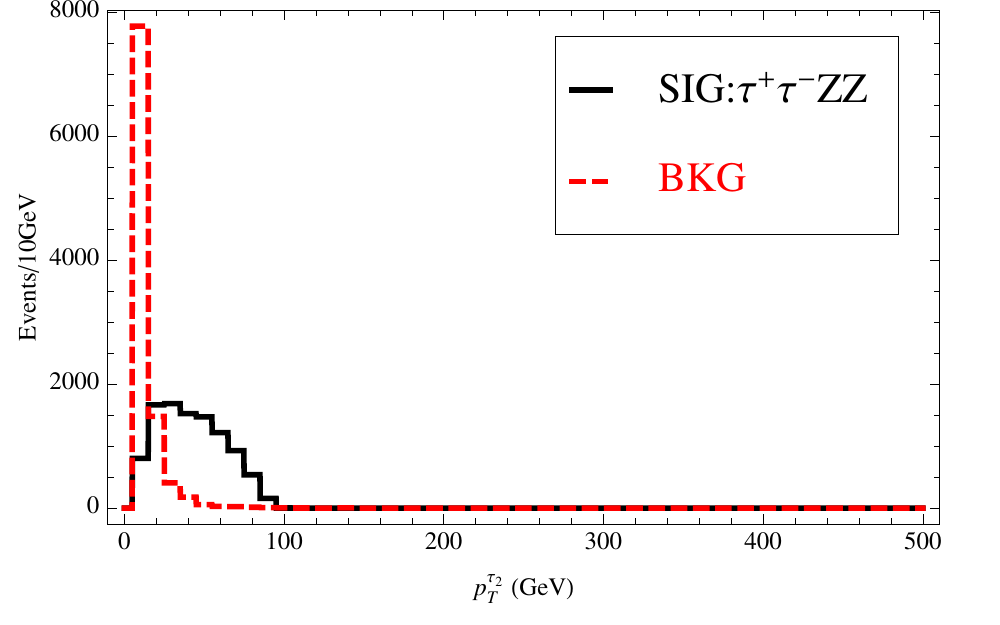}
\includegraphics[width=7.5cm]{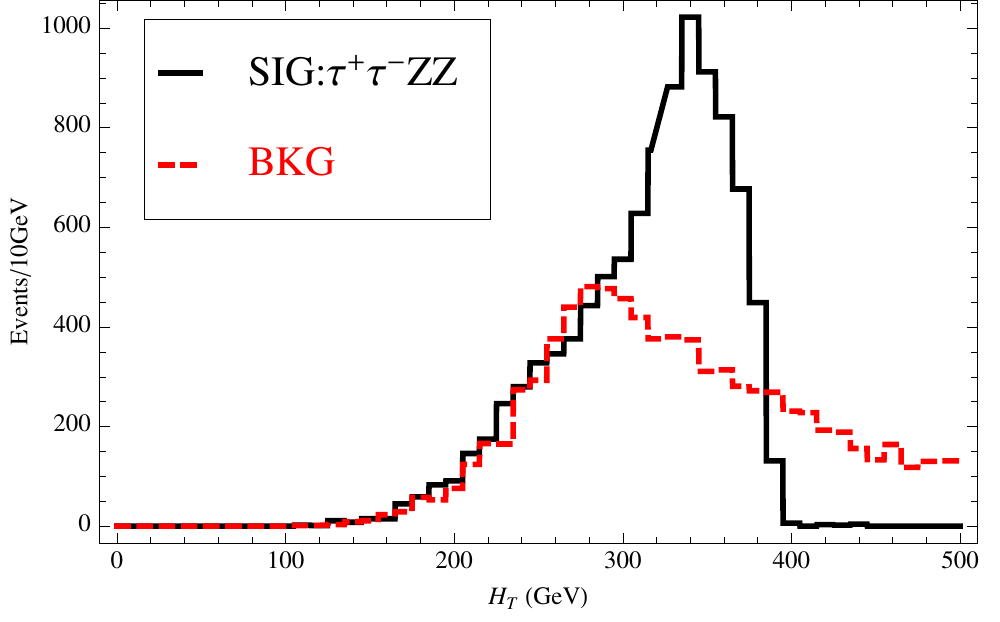}
\caption{The $p_T$ and $H_T$ distribution for signal $\tau^-\tau^+ ZZ$ and the corresponding  background. The dashed red line is background and the solid black line is signal.}\label{fig:2tau2Z}
\end{figure}

In this section, we assume the signal process is $pp\rightarrow S\rightarrow HH\rightarrow(\tau^-\tau^+)(ZZ^{*})$ with the $Z$ boson decay leptonically. At the same time, the background contains $pp\rightarrow \tau^-\tau^+(Zl^+l^-)$. In order to choose the suitable cut conditions, we give the $p_T$ and $H_T$ distribution in Fig.\ref{fig:2tau2Z}.

With the distribution shown in Fig.\ref{fig:2tau2Z}, we choose the cut conditions below,
\begin{eqnarray}
\begin{split}
&&p^{\tau_1}_T\geq 80GeV, p^{\tau_2}_T\geq 20GeV, H_T\leq 400GeV.
\end{split}
\end{eqnarray}

We also impose the invariant mass cut conditions to further suppress the background.

\begin{eqnarray}
\begin{split}
&&m_{4l}, m_{\tau^-\tau^+} \in [115GeV,135GeV], m_{4l2\tau} \in [360GeV,440GeV]
\end{split}
\end{eqnarray}

After above cuts, the event number for signal and background are shown in Table.\ref{tab:2tau2Z}. This channel is very clean due to the six leptons in the final states, the electroweak background is quite small. The prompt background has a very small cross-section comparing with signal. The non-prompt background from $\tau$ misidentification is also smaller than signal after applying the $\tau$ tagging efficiency. Though clean background it has, the signal statistics is low with about 70 for $1000fb^{-1}$ integrated luminosity. Such signal is quite rare,  but has very small background and worth looking at.

 \begin{table}[tbp]
\centering
\begin{tabular}{lcc}
\hline
event number &signal &BKG($l^-l^+l^-l^+\tau^-\tau^+$)\\ \hline
cross-section($pb$)&$3.4 \times 10^{-4}$&  $4.829 \times 10^{-6}$ \\
cut efficiency &  0.8177 &  $6.7\times10^{-5}$ \\
$\tau_h$ tag   &      $(50\%)^{2}$      &  $(50\%)^{2}$   \\
event number &  70  &  $8\times10^{-5}$\\
$\frac{S}{B}$ ratio& $9\times10^5$ \\ \hline
\end{tabular}
\caption{The cut flow table for two Higgs decay to $\tau^-\tau^+ ZZ$. In the cross-section for background, we may have added prior $p_T$ requirement in the Madgraph to generate events more efficiently.}\label{tab:2tau2Z}
\end{table}

\section{Discussion and Conclusion}\label{sec:conclusion}
We have discussed the resonant production of heavy scalar $S$, which subsequently decay into two Higgs $H$. Since the decay channels of $H$ have different properties, we list all the combinations of final state of two Higgs decay and explore the possible signal in the LHC14 with integrated luminosity$mathcal{L}=1000fb^{-1}$. Generally, we find that with the help of invariant mass information from H and S, very good signal to background ratio $S/B$ can be archived in many decay channels.
We summarize the discovery potential for different combinations of decay channels in the Table.\ref{tab:discovery}.

For one Higgs decay to $b\bar{b}$, the general properties are the signal benefits from the large decay BR, but suffer from large QCD background. The two Higgs both decay to $b\bar{b}$ are very hard to discover due to QCD background. It needs leptons or photons in the final state to suppress the QCD background. For the other Higgs not decay to $b\bar{b}$ but others, the signals have very good $S/B$ ratio, usually larger than 100 while the signal statistics are good at the same time, with event numbers from several hundreds to tens of thousands.

For one Higgs decay to $\gamma\gamma$, the general properties are the signal suffers from the  very small decay BR, but benefits from very small QCD background. The prompt electroweak background in this channel is usually very small comparing with signal. But one should be cautious about the non-prompt photons from jet misidentification. For the two Higgs both decay to $\gamma\gamma$, the signal statistics is quite low, only about 35 and we also should note that non-prompt photons may further reduce the $S/B$ ratio that this channel is not very good. For the other Higgs decay to $ZZ^{*}$, it is also not a good signal to search due to even lower statistics of about 9. Despite of these two cases, for the other Higgs decay to other particles, the signal has good statistics from hundreds of events to thousands. The $S/B$ ratio is also high, larger than several hundreds.

For one Higgs decay to $ZZ^{*}$, the general properties are the signal suffers from small leptonic decay BR, sometimes even lower than $\gamma\gamma$, but it benefits from multi-lepton final states which greatly suppress the large QCD background. This channel usually needs the other Higgs decay to a channel which has large BR. For two Higgs both decay to $ZZ^{*}$, one should require one pair of $Z$ boson decay hadronically to avoid the low statistics. Even in this case, the event number is about 46 which is not large but the background is extremely small. For the other Higgs decay to other particles, $b$ quark and $W$ boson, the signals generally has event numbers of several hundreds and very good $S/B$ ratio. For the other Higgs decay to tau leptons, the signal event numbers is lower, of about 70, but this is a six lepton final states which is very clean.

For one Higgs decay to $WW^{*}$, the general properties are the signal should require $W$ decay leptonically to suppress the large QCD background, while due to the large leptonic decay BR of $W$ and $WW^{*}$ BR of Higgs the signal usually has very good statistics. But one should note that the $W$ decay leptonically has missing energy, which in the real search one needs to reconstruct the MET by the on-shell mass requirement which may reduce the signal and increase the background. No matter what does the other Higgs decay to, the signal usually has good statistics larger than hundreds of event number and good $S/B$ ratio.

For one Higgs decay to $\tau^{-}\tau^{+}$, the general properties are the signal benefits from the modest decay BR while can suppress large QCD background from lepton final states. For the $\tau$ final states, one should note that the jet has probability to misidentify the hadronic $\tau$, which will increase the background. We have done an estimate of such background and found that they usually much smaller than signal, though they reduce the $S/B$ ratio to about 100. For the other Higgs decay other than $ZZ^{*}$, the signals usually have very good statistics, larger than several hundreds events, while the $S/B$ ratios are quite good usually larger than 100. For the other Higgs decay to $ZZ^{*}$, one should note that the event number is about 70 but this channel is quite clean.

Last but not least, we should emphasize that the simulation in this paper is quite rough and further detailed simulation should put the conclusions on more solid ground.

\section*{Acknowledgment}

This work was supported in part by the Natural Science Foundation of China (Nos. 11075003, 11135003 and 11375014).

\bibliography{referencelist}

\end{document}